\documentclass[superscriptaddress,prx,nofootinbib,notitlepage]{revtex4-2}
\usepackage{graphicx}
\usepackage{amssymb,amsmath,amsthm,amsfonts}
\usepackage{thmtools}
\usepackage{mathtools}
\mathtoolsset{centercolon}
\usepackage{thm-restate} 
\usepackage{amsmath}
\usepackage{amssymb}
\usepackage{comment}
\usepackage{placeins}
\usepackage{multirow}
\usepackage[caption=false]{subfig}
\usepackage[colorlinks]{hyperref}
\usepackage{tikz}
\usepackage{dcolumn}
\usepackage{tabularx}
\usetikzlibrary{calc,backgrounds}

\usepackage{pgffor}
\usepackage{url}
\usepackage{hypcap}
\usepackage{dsfont}
\usepackage{algorithm}
\usepackage{algorithmic}
\usepackage{soul}
\usepackage{diagbox}

\usepackage[version=4]{mhchem} 
\usepackage{siunitx} 
\DeclareSIUnit{\atom}{atom}
\DeclareSIUnit{\lno}{f.u.}

\newcommand{\eq}[1]{Eq.~\hyperref[eq:#1]{(\ref*{eq:#1})}}
\renewcommand{\sec}[1]{\hyperref[sec:#1]{Section~\ref*{sec:#1}}}
\newcommand{\app}[1]{\hyperref[app:#1]{Appendix~\ref*{app:#1}}}
\newcommand{\tab}[1]{\hyperref[tab:#1]{Table~\ref*{tab:#1}}}
\newcommand{\fig}[1]{\hyperref[fig:#1]{Figure~\ref*{fig:#1}}}
\newcommand{\figa}[2]{\hyperref[fig:#1]{Figure~\ref*{fig:#1}#2}}
\newcommand{\figx}[2]{\hyperref[fig:#1]{Figure~\ref*{fig:#1}(#2)}}
\newcommand{\thm}[1]{\hyperref[thm:#1]{Theorem~\ref*{thm:#1}}}
\newcommand{\lem}[1]{\hyperref[lem:#1]{Lemma~\ref*{lem:#1}}}
\newcommand{\cor}[1]{\hyperref[cor:#1]{Corollary~\ref*{cor:#1}}}
\newcommand{\defn}[1]{\hyperref[def:#1]{Definition~\ref*{def:#1}}}
\newcommand{\alg}[1]{\hyperref[alg:#1]{Algorithm~\ref*{alg:#1}}}

\newcommand{\be}{\begin{equation}}
\newcommand{\ee}{\end{equation}}
\newcommand{\ba}{\begin{eqnarray}}
\newcommand{\ea}{\end{eqnarray}}

\usepackage{color}
\usepackage{xcolor}

\makeatletter
\renewcommand*\env@matrix[1][\arraystretch]{%
	\edef\arraystretch{#1}%
	\hskip -\arraycolsep
	\let\@ifnextchar\new@ifnextchar
	\array{*\c@MaxMatrixCols c}}
\makeatother

\makeatletter

\newcommand*{\@rowstyle}{}

\newcommand*{\rowstyle}[1]{
  \gdef\@rowstyle{#1}%
  \@rowstyle\ignorespaces%
}

\newcolumntype{=}{
  >{\gdef\@rowstyle{}}%
}

\newcolumntype{+}{
  >{\@rowstyle}%
}

\makeatother

\makeatletter
\def\DC@endright{$\hfil\egroup\@dcolcolor\box\z@\box\tw@\dcolreset}
\def\dcolcolor#1{\gdef\@dcolcolor{\color{#1}}}
\def\dcolreset{\dcolcolor{black}}
\dcolcolor{black}
\makeatother

\begin{document}

\newcommand{\Google}{\affiliation{%
Google Quantum AI, Venice, CA 90291, United States}}

\newcommand{\QSimulate}{\affiliation{%
Quantum Simulation Technologies, Inc., Boston, 02135, United States}}

\newcommand{\Columbia}{\affiliation{%
Department of Chemistry, Columbia University, USA}}
\newcommand{\Toronto}{\affiliation{%
Department of Computer Science, University of Toronto, Canada}}

\newcommand{\Macquarie}{\affiliation{Department of Physics and Astronomy, Macquarie University, Sydney, NSW, Australia}}

\newcommand{\BI}{\affiliation{%
Quantum Lab, Boehringer Ingelheim, 55218 Ingelheim am Rhein, Germany}}

\newcommand{\BIMedChem}{\affiliation{%
Boehringer Ingelheim Pharma GmbH \& Co KG, Birkendorfer Strasse 65, 88397 Biberach, Germany}}

\newcommand{\UIBK}{\affiliation{Department of General, Inorganic and Theoretical Chemistry,
University of Innsbruck,
6020 Innsbruck, Austria}}

\title{Reliably assessing the electronic structure of cytochrome P450\\ on today's classical computers and tomorrow's quantum computers}

\author{Joshua J.~Goings}
\thanks{These authors contributed equally}
\Google

\author{Alec White}
\thanks{These authors contributed equally}
\QSimulate

\author{Joonho Lee}
\Google
\affiliation{Department of Chemistry, Columbia University, New York, New York 10027, USA}

\author{Christofer S.~Tautermann}
\BIMedChem \UIBK

\author{Matthias Degroote}
\BI

\author{Craig Gidney}
\Google

\author{Toru Shiozaki}
\QSimulate

\author{Ryan Babbush}
\email{ryanbabbush@gmail.com}
\Google

\author{Nicholas C.~Rubin}
\email{rubinnc0@gmail.com}
\Google

\begin{abstract}
An accurate assessment of how quantum computers can be used for chemical simulation, especially their potential computational advantages, provides important context on how to deploy these future devices. In order to perform this assessment reliably, quantum resource estimates must be coupled with classical simulations attempting to answer relevant chemical questions and to define the classical simulation frontier. Herein, we explore the quantum and classical resources required to assess the electronic structure of cytochrome P450 enzymes (CYPs) and thus define a classical-quantum advantage boundary.  This is accomplished by analyzing the convergence of DMRG+NEVPT2 and coupled cluster singles doubles with non-iterative triples (CCSD(T)) calculations for spin-gaps in models of the CYP catalytic cycle that indicate multireference character. 
The quantum resources required to perform phase estimation using qubitized quantum walks are calculated for the same systems. Compilation into the surface-code provides runtime estimates to compare directly to DMRG runtimes and to evaluate potential quantum advantage. Both classical and quantum resource estimates suggest that simulation of CYP models at scales large enough to balance dynamic and multiconfigurational electron correlation has the potential to be a quantum advantage problem and emphasizes the important interplay between classical simulations and quantum algorithms development for chemical simulation.
\end{abstract}

\maketitle

\section{Introduction}
Chemical simulation is among the most promising applications of quantum computers. Despite this, it remains a challenge to accurately assess and identify chemical problems for which one can reasonably expect future quantum computational advantage. This problem is challenging for many reasons, but two key difficulties emerge when demarcating the boundary of quantum computational advantage. First, given the myriad conventional polynomial scaling electronic structure methods, it is difficult to find chemical problems which will not yield to at least one classical method. For any claim that a problem is classically difficult—or even impossible—there is no guarantee that the claim will not be challenged by a new method at some time later. The second difficulty is that quantum algorithms for chemistry are still an active area of development, so estimates of the resources required to compile and run experiments on quantum computers will continue to evolve. Until the development of a truly scalable and fault-tolerant quantum computer, resource estimates of the most promising quantum algorithms are limited to rigorous calculations of prefactors yielding run-time upper bounds. Moreover, a chemical problem may prove to be prohibitive on \emph{either} a quantum or a classical computer.
 
Thus, any attempts to determine the boundary of quantum computational advantage must involve high accuracy classical quantum chemistry simulations along with a detailed resource estimation of quantum algorithms and the cost of measuring chemically relevant observables. Ideally, quantum advantage is defined within a realistic model of chemistry and is associated with a computation that answers a typical chemistry question. In this work, we articulate the nuances in describing this boundary concretely by focusing on the quantum and classical resources required to reliably simulate the active space of a biologically important enzyme.  We simulate the active space of cytochrome P450 (CYP) mimics with a variety of classical electronic structure methods to assess the degree of strong correlation and what would be required to: a) evaluate the chemical mechanism of reactivity for CYPs and b) the spin state ordering of reactive intermediates of the catalytic cycle of CYPs which are necessary for a correct description of energy barriers. To assess the quantum cost we evaluate run times and logical qubit requirements required to implement phase estimation within the surface code error correction scheme.  

\textcolor{black}{We focus on costing out these two questions as they are representative of the types of chemical questions one could ask about a realistic system which highlight the difficulties of application of quantum and classical algorithms in chemical science. To this end, we use several systematically improvable classical electronic structure methods (discussed below) to scope out the limits of classical algorithms to obtain reliable chemical energetics. Having shown that classical calculations will be cost-prohibitive, we compute the resource estimates for quantum phase estimation algorithms, which would be required for efficient and reliable computation of the energetics in CYP mimics. What emerges from these detailed accountings of cost is that despite a potential exponential simulation time advantage for quantum computers, direct simulation of a large enough system to reliably account for dynamic correlation may be beyond the reach of classical and quantum devices.  This suggests room for further development of quantum algorithms.}

To quantify the cost, and therefore the limits, of classical electronic structure calculations we focus on families of methods which are systematically improvable. Such methods have the property that one can estimate convergence of any given property towards the exact result, and this is the basis of truly predictive computation. One such family of methods is the coupled cluster (CC) hierarchy \cite{Coester1958,Cizek1966,Paldus1972,Cizek1980,Bartlett1981,Purvis1982,Bartlett2007} which, for most systems, provides results with controllable error at a high, but polynomial, cost. For systems with significant multireference character, coupled cluster methods will fail, and one must turn to multireference methods usually based upon an exact or near-exact solution within an active space of orbitals. In theory, such methods can be systematically improved by increasing the size of the active space, but in practice the high cost often necessitates small active spaces or aggressive truncations to the molecular model for which the error is difficult to control. For small active spaces, the full configuration interaction (FCI) calculation can be performed directly, while for larger active spaces one must turn to approximate methods such as the density matrix renormalization group (DMRG) \cite{White1992,Green1999,Chan2002,Chan2009}, full configuration interaction Monte Carlo (FCIQMC) \cite{Booth2009}, or some variant of selected configuration interaction method \cite{Huron1973, Holmes2016, Schriber2016, Tubman2016}. Though research into these methods has enabled large active space calculations in recent years, medium- to large-sized molecular systems will still require some treatment of electron correlation outside of the active space to obtain qualitatively correct results. Progress has been made with various flavors of multireference perturbation theory \cite{Andersson1990,Angeli2001} and multireference coupled cluster theory \cite{Lyakh2012}, but a balanced description of electron correlation outside of the active space remains a major challenge.  

The quantum algorithm we choose to compare against is phase estimation which allows one to sample in the eigenbasis of a given Hamiltonian. Assuming one can prepare an initial state with sufficient overlap with the ground state, phase estimation can be used to estimate the ground state energy of a chemical system. Though there are a variety of quantum algorithms for acquiring different types of chemical observables, energy computations provide direct comparison to address the aforementioned problem of spin-gap estimation with single-point energy calculations.  
Improvements to the algorithm have brought costs down from prohibitively high runtimes to scalings that now scale linearly in inverse precision and the square root of the basis-independent information content of the Hamiltonian~\cite{PRXQuantum.2.030305}. However, asymptotic scalings are not enough to define a quantum advantage boundary. Prefactor estimates along with compilation considerations have been taken into account to provide an upper bound to a real-time estimate for the runtime of a quantum algorithm subject to the assumption of high initial overlap.  To make a direct comparison to classical simulation costs in active spaces of chemical systems, we perform a quantum resource estimation assuming phase estimation of the qubitized quantum walk operator~\cite{PRXQuantum.2.030305}.  
Qubitization strongly depends on the type of tensor factorization one uses for the two-electron integrals in the standard electronic structure Hamiltonian.  We study this dependency for three different factorizations and demonstrate that tensor hypercontraction leads to substantially lower runtimes in CYP systems.  Finally, we analyze tradeoffs in qubit count and number of Toffoli factories to define a quantum advantage frontier for a variety of hardware configurations.

To compare cost of classical and quantum computation, we focus on models of the active site of CYP proteins. Compared to exotic systems, like the FeMo cofactor of nitrogenase (FeMoco)~\cite{PRXQuantum.2.030305}, that are usually used for quantum resource estimates, the active sites of CYP proteins are more representative of the typical systems which can benefit from chemical simulation. The superfamily of CYPs are membrane-bound heme-containing enzymes which function mostly as monooxygenases. In the human genome 57 CYP isoforms are encoded, moreover, it is the largest family of hemoproteins known, with more than 300,000 members throughout all organisms\cite{Nelson:2018}. The major role of CYPs lies in the detoxification of organisms. The most common detoxification mechanism involves a single oxygen insertion into C-H bonds of CYP substrates, thereby generating a hydroxy-group, which enables further metabolism. The oxidation by CYPs is also the most common metabolization mechanism for drugs in humans, where more than 70\% of all drugs are metabolized by just two CYP isoforms (CYP 3A4 and CYP 2D6) \cite{Guengerich:2008}. In the case of CYP 3A4, which is known to roughly metabolize 50\% of all marketed drugs \cite{Liu:2007}, more than one substrate molecule may be accommodated in the active site \cite{Boobis:2009}.
    
\begin{figure}[ht]
\centering
\includegraphics[width=0.75\textwidth]{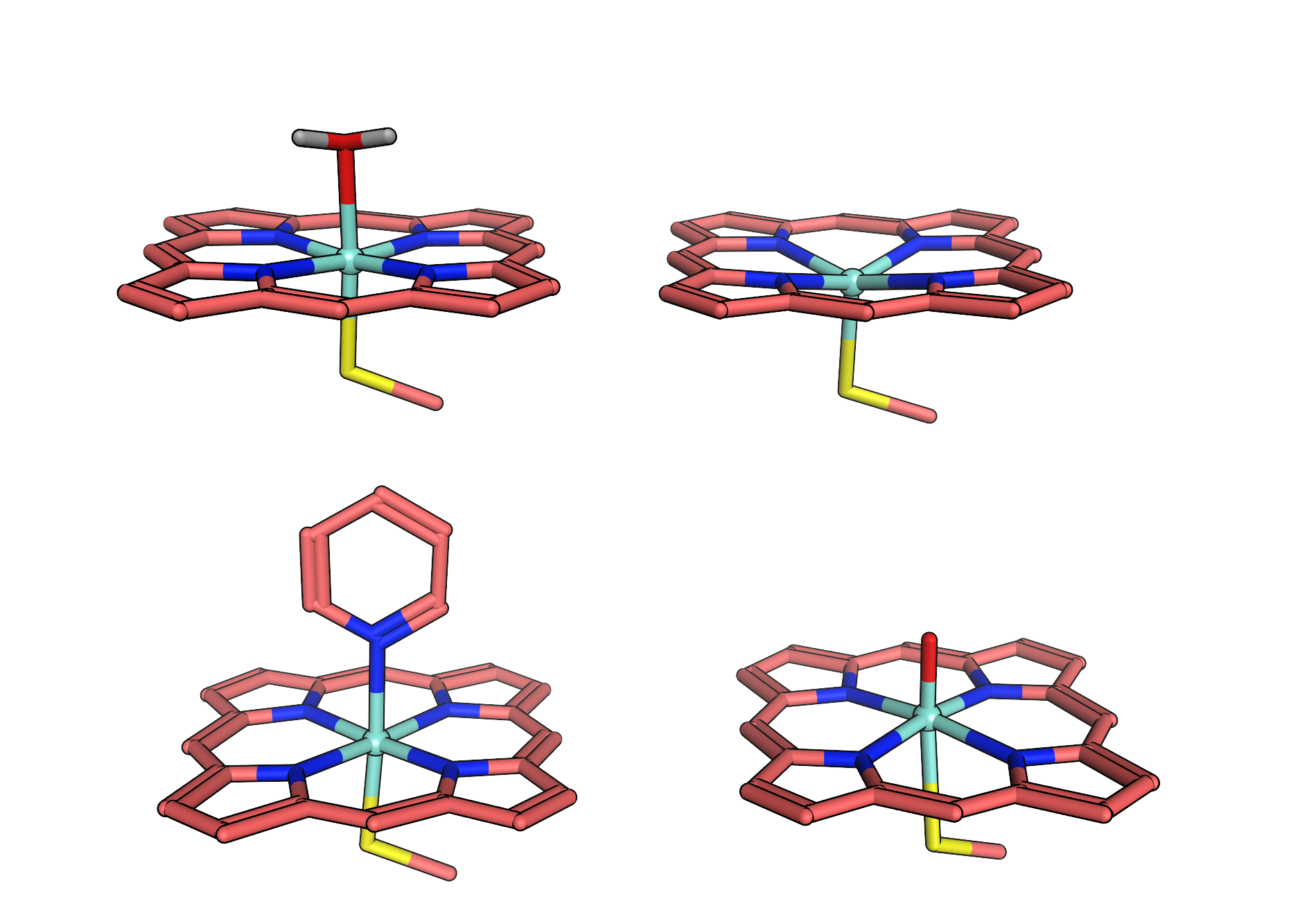}
\caption{\label{fig:model} Model systems employed in this study: \emph{top left}: the resting state with water bound to heme, \emph{top right}: the pentacoordinated ``empty'' state, \emph{bottom left}: pyridine inhibitor bound model complex, \emph{bottom right}: Cpd I. To improve clarity all non-polar hydrogen atoms are hidden. 
}
\end{figure}

The oxidation by CYPs is a multi-step catalytic cycle, in  detail shown in Appendix~\ref{app:P450} and Figure~\ref{fig:catal_cycle}, involving at least eight intermediates \cite{Shaik:2010}, consuming one molecule of O\textsubscript{2}, two electrons, and two protons to achieve the monooxygenation of a substrate. The catalytically active species, compound I (Cpd I) \cite{Shaik:2010}, is thought to be a polyradical neutral species involving an iron porphyrin ring coordinating to atomic oxygen and thiolate from cysteine. The spin state energetic orderings of Cpd I are still a matter of debate, where nearly-degenerate doublet and quartet states are postulated \cite{Shaik:2010}. The spin states are sensitive to the protein environment and Cpd I is often referred to as a ``chameleon species", as it changes its spin nature depending on subtle structural changes in the protein. A complete theoretical description of this catalytic cycle is challenging because incorrect results for a single intermediate can qualitatively change the chemistry.  If a single intermediate has an electronic structure demanding a more accurate treatment, then for consistency all intermediates should be calculated at this level of theory. To portray the classical and quantum costs along the CYP catalytic cycle, we decided to employ model systems of the water-bound resting state, the pentacoordinate ``empty'' state, and Cpd I. Additionally, due to the high pharmaceutical relevance, we also include a model system corresponding to an inhibitor-bound active site of P450, where a pyridine is bound to the heme-iron, such as shown in Figure~\ref{fig:model}.

In this work we analyze the model compounds with state-of-the-art classical electronic structure methods, estimate the quantum resources required for different-sized active space Hamiltonians, and provide a classical characterization of the electron correlation in these systems. In Section~\ref{sec:estruct} we describe the series of active space models studied and DMRG calculations with an $n$-electron valence state perturbation theory (NEVPT2) correction to determine accurate spin-gaps and discuss potential hero calculations that can be performed to resolve the spin-gap problem.  In Section~\ref{sec:quantum_comp_resource_est} we provide quantum resource estimates and runtimes after compiling to a surface-code quantum error correction scheme. With full compilations to realistic hardware configurations we compare runtimes from DMRG and the quantum computer for the task of simulating the ground state energy of the active space models. 
In Section~\ref{sec:justifications} we characterize the electronic structure of four model CYP compounds and demonstrate that the empty, inhibited, and resting states can be characterized by single-reference electronic structure methods while Cpd I exhibits some multiconfigurational character. Specifically, we make the distinction that traditional metrics for strong correlation, $\mathrm{max}(|t_{1}|)$-diagnostic and spin-contamination, are corrected by including dynamic correlation and thus the three aforementioned compounds are classified as ``artificially'' symmetry broken. 
We close with a discussion of future research directions and chemically relevant observables for characterizing CYP with quantum or classical computation.  In addition to providing a robust protocol for comparing the costs of quantum and classical calculations on realistic systems, our results suggest a substantial quantum advantage in those cases that require high-accuracy results.

\section{Classical calculations of electron correlation in p450 models}\label{sec:estruct}
The energetics, dominant electronic structure features, and spin gaps have been subjects of many quantum chemistry calculations on various CYP isoforms~\cite{Shaik:2010, Chen2011, Chen2011a, lee2020utilizing, Phung:2019}, ranging from full space DFT calculations~\cite{groenhof2005electronic, matsuzawa1995density, rovira1997equilibrium, radon2014spin} to active space models with dynamic correlation corrections~\cite{vancoillie2011multiconfigurational, phung2016cumulant, zhou2019multiconfiguration}.
The model systems in this work were derived from experimental X-ray structures of CYP3A4 by removing all non-iron coordinating entities of the protein and the solvent. Further details on how the geometries were determined can be found in Appendix~\ref{app:geometries_from_xray}. In the following section we describe the construction of a hierarchy of active spaces and use CCSD(T) and DMRG with NEVPT2 corrections to characterize the electronic structure at various experimentally motivated spin states.  The DMRG timings and accuracy are used to provide context for the quantum resource estimates and ultimately motivate a potential quantum advantage boundary. The computational details for all calculations can be found in Appendix~\ref{app:software_implementation_details}.

\subsection{Active space selection}

To create the active space models, the orbitals of the high-spin ROHF state in the cc-pVDZ basis~\cite{Dunning1989} were localized with the Pipek-Mezey localization scheme~\cite{pipek1989} to yield a set of local orbitals for each compound. We constructed active spaces of increasing size in a hierarchical manner by starting with the five singly-occupied orbitals of the high spin reference (A) and then adding orbitals from the occupied and virtual spaces as summarized in Table~\ref{tab:active_space}. For heme-iron systems the computed spin gap will depend strongly on the choice of active space\cite{Radon2007, Radon2008,Vancoillie2010,Radon2010,Phung2016,Pierloot2017,LiManni2018, Weser2021}. While most previous studies seek the most efficient possible active space of a given size, our strategy is designed to yield a balanced hierarchy of active spaces that will facilitate analysis of computational cost and convergence to the exact limit.

This process was repeated for each of the four compounds to yield a hierarchy of active spaces (A, B, C, D, E, F, G, X), each a superset of the previous, all with roughly the same filling fraction. The numbers of orbitals and electrons in each active space are tabulated in Appendix~\ref{app:active_space_orb_n}.

\begin{table}[h]
\caption{A summary of our strategy for constructing the hierarchy of active spaces used in this work. Each active space is formed by adding the listed orbitals to the active space on the preceding line beginning with the `A' active space consisting of only the singly occupied orbitals. We show the size of the active space, in number of orbitals, for our model of Cpd I in the final column.}
\label{tab:active_space}
\begin{ruledtabular}
\begin{tabular}{cllc}
    name & \multicolumn{1}{c}{occupied orbitals} & \multicolumn{1}{c}{virtual orbitals} & size (Cpd I)\\ \hline
    A & Singly-occupied orbitals & - & 5\\
    B & Iron 3d & Iron 4s/iron-nitrogen anti-bonding & 8\\
    C & Iron-axial ligand(s) bonding& Iron 4p/d'/iron-axial ligand anti-bonding & 15\\
    D & Heme iron-nitrogen bonding& Iron 4p/d'/iron-nitrogen anti-bonding & 23\\
    E & Heme $\pi$ bonding (localized around N atoms)& Heme $\pi$ anti-bonding (localized around N atoms)& 31\\
    F & Heme $\pi$ bonding (localized around C atoms)& Heme $\pi$ anti-bonding (localized around C atoms)& 41\\
    G & Cysteine carbon-sulfur bonding& Cysteine carbon-sulfur anti-bonding& 43\\
    X & Heme carbon-nitrogen bonding& Heme carbon-nitrogen anti-bonding&  58
\end{tabular}
\end{ruledtabular}
\end{table}

\subsection{Spin-state ordering from DMRG+NEVPT2}

\begin{figure}[ht]
\centering

\includegraphics[width=0.3\textwidth]{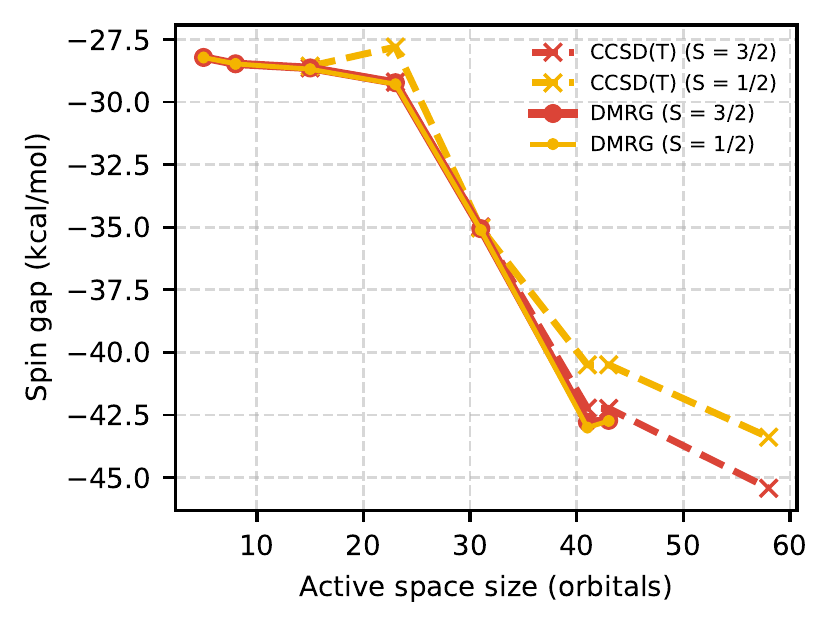}
\caption{\label{fig:cpd1_energies} 
Energy differences between the high-spin sextet (S=5/2) and low-spin doublet (S=1/2) and quartet (S=3/2) states of Cpd I. 
Only for the doublet is there significant disagreement between CCSD(T) and DMRG energies within the active space. DMRG consistently predicts the doublet and quartet state to be nearly degenerate.}
\end{figure}

\begin{figure}[ht]
\centering
\includegraphics[width=0.3\textwidth]{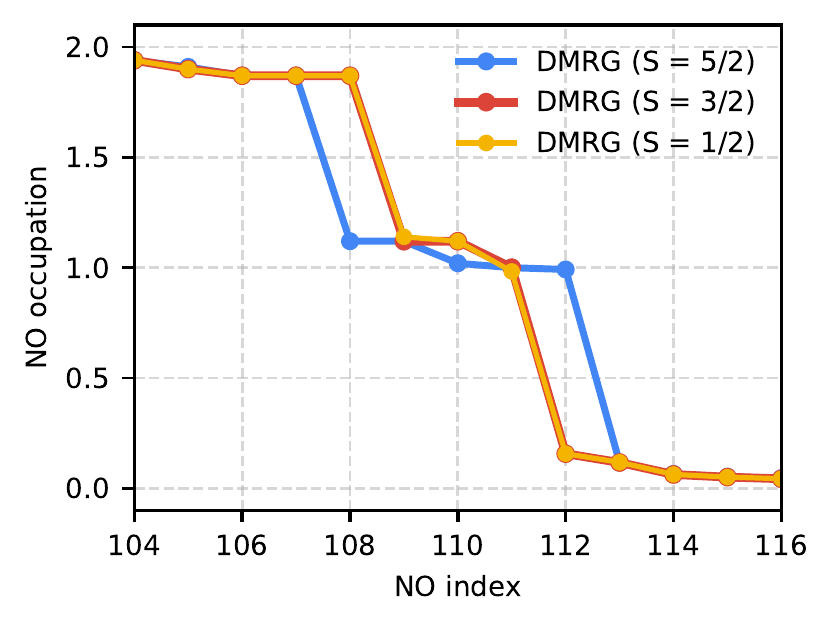}
\caption{\label{fig:cpd1_NO} Natural orbital occupation numbers from the DMRG density matrix in the 41 orbital G active space for the sextet (S=5/2), quartet (S=3/2), and doublet (S=1/2) states. Note that the doublet clearly has three singly occupied orbitals.}
\end{figure}

In all active spaces, we found the sextet to be much higher in energy than the nearly-degenerate doublet and quartet states. This is largely consistent with previous calculations on models of the system \cite{Shaik:2010, Chen2011, Chen2011a}. In Figure~\ref{fig:cpd1_energies} we show the DMRG and CCSD(T) spin gaps within the specified active space. CCSD(T) and DMRG agree to within 0.1 kcal/mol for the sextet and quartet states indicating single-reference character. However, for the doublet state, the DMRG energy is consistently much lower than the CCSD(T) energy. The DMRG natural orbital (NO) occupation numbers shown in Figure~\ref{fig:cpd1_NO} reveal that the doublet state has three open-shell natural orbitals. This state is not well-described in traditional coupled cluster theory, and accurate computation therefore requires a method, such as DMRG, that is capable of treating systems with multiconfigurational character.

\begin{figure}[ht]
\centering
\includegraphics[width=0.9\textwidth]{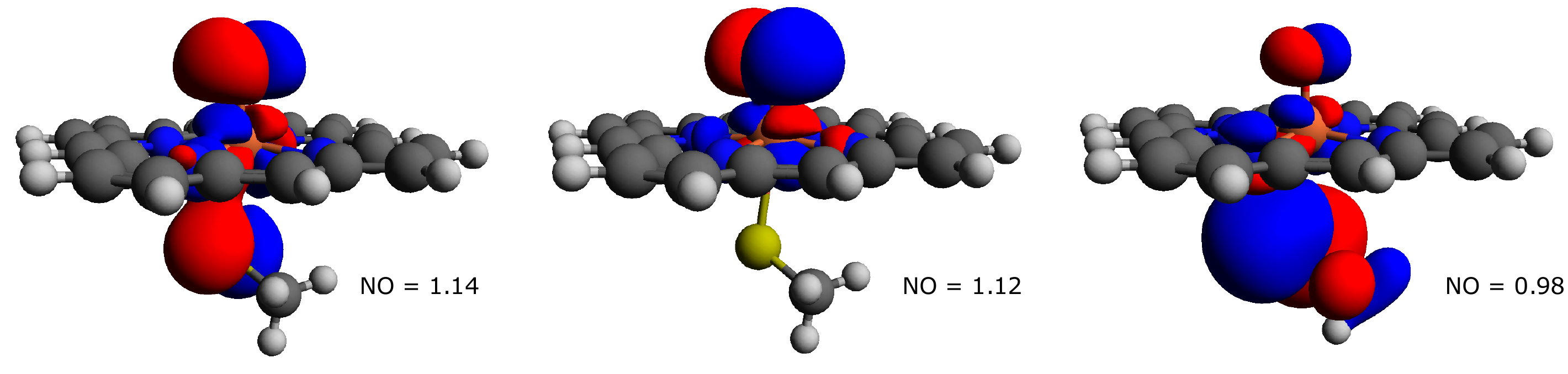}
\caption{\label{fig:cpd1_F_NO}Visualization of the three singly occupied natural orbitals of the doublet state of Cpd I.}
\end{figure}

Our DMRG results for the G active space are the largest DMRG calculations yet performed on any model of Cpd I. We find the doublet state to be lower than the quartet by only 0.02 kcal/mol. Thus, from these estimates both the doublet and quartet states would be expected to be populated at room temperature. The sextet is 42.7 kcal/mol higher in energy than the two low spin states. This is qualitatively consistent with most past calculations on this system.~\cite{Antony1997, Green1999, Shaik:2010, Chen2011, Chen2011a} We show the 3 singly occupied NOs in Figure~\ref{fig:cpd1_F_NO}  from which we can see that there are two Fe-O orbitals with $\pi_{xz}^{\ast}$ and $\pi_{yz}^{\ast}$ character and a third sulfur non-bonding orbital. In this case the sulfur non-bonding orbital is mixed with one of the Fe-O $\pi^{\ast}$ orbitals. This is consistent with studies that use S-Me or the full cysteine ligand in the gas phase \cite{Antony1997, Green1999}. In studies that either use S-H in place of the cysteine ligand \cite{Ogliaro2000, Radon2007} or include the protein environment in some way \cite{Schoneboom2005, Altun2008}, the third orbital is found to be mixed with a heme $a_u$ ($\pi^{\ast}$) orbital.

The near-degeneracy of the doublet and quartet states makes prediction of the lowest energy spin state very difficult for this system. In particular, including the remaining dynamic correlation from outside of the active space can change this picture qualitatively (see the results in Appendix~\ref{app:nevpt2}). While NEVPT2 can provide a good estimate of the relatively large dynamic correlation energy, it is not accurate enough to resolve the doublet and quartet states. In this system, the addition of the NEVPT2 correlation energy can even shift the relative energy of the sextet such that it becomes the lowest spin state in some of the active spaces. Without larger, prohibitively expensive calculations, one cannot distinguish a real effect from an artifact of the various approximations. Furthermore, details of the protein model, dielectric environment, and cysteine ligand model can easily shift the energies by enough to change the qualitative result~\cite{shaik2005theoretical}. Together, these sources of uncertainty indicate that reliable identification of the lowest spin state of Cpd I as it appears in most experiments is not feasible. However, unambiguous identification of the lowest spin state in our model compound could be possible with a DMRG calculation in a very large active space, such as our `X' active space, followed by a NEVPT2 correction for the remaining dynamic correlation. As the active space is made larger, the NEVPT2 correction becomes smaller and more reliable, though a DMRG-NEVPT2 calculation of this size is beyond our current capability. It is interesting to note that in a system like this, a very large active space is required not because there are a large number of strongly correlated orbitals, but rather because this is the only means to obtain a balanced combination of static and dynamic correlation for different spin states.

\subsection{Cost estimates for classical computation}
The computational cost of DMRG calculations depends on the bond dimension, $M$, and number of active orbitals, $k$. The bond dimension, $M$, is an adjustable parameter that controls the quality of the calculation, and convergence of the energy with respect to $M$ must be carefully monitored to ensure accurate results. The theoretical asymptotic scaling of computational resources required for a DMRG calculation with a given bond dimension and number of active orbitals has been discussed elsewhere \cite{olivares-amaya2015}. The scaling of the CPU time is $O(k^3M^3)$, and the memory and disk requirements scale as $O(k^2M^2)$ and $O(k^3M^2)$ respectively. Given the theoretical scaling, we can estimate the computational cost of a calculation in the 58 orbital X active space. These estimates are shown in Table~\ref{tab:dmrg_resources}. For the StackBlock program, which implements both shared-memory and distributed-memory parallelism, it may be possible to perform an $M = 3000$ calculation on the 58 orbital X active space with a significant investment of computational resources over a period of approximately one month of wall time. However, there is no guarantee that a bond dimension of $M = 3000$ will be sufficient and it is likely that calculations of even higher bond dimension are required. 
\begin{table}[h]
\centering
\begin{minipage}{0.75\textwidth}
\caption{Actual resources required for DMRG calculations in the 43 orbital G active space and estimated resources needed for DMRG calculations on the 58 orbital X active space.}
\label{tab:dmrg_resources}
\begin{ruledtabular}
\begin{tabularx}{0.7\textwidth}{cccc}
     & G ($M = 1500$)& X ($M = 1500$)& X ($M = 3000$)\\ \hline
     CPU time (hrs) & 1800& 4570& 36564\\
     Memory (Gb) & 48& 87& 348\\
     Disk (Gb) & 235& 572& 2288\\
\end{tabularx}
\end{ruledtabular}
\end{minipage}
\end{table}

For systems like our model of Cpd I which do not have a quasi one-dimensional structure the bond dimension required to converge a DMRG calculation to the required precision will grow in such a way so as to make the overall scaling of the method weakly exponential. In practice, this means that active spaces larger than the X active space described here quickly become intractable.

\section{Quantum Computing Resource Estimates}\label{sec:quantum_comp_resource_est}
In this section we perform a detailed accounting of the space and time complexity for sampling from the eigenbasis of the active space Hamiltonians within the context of error corrected quantum computers. The space resources we consider are the total number of logical qubits and physical qubits required to perform phase estimation.  For time complexity we focus on the number of Toffoli gates which are the rate-limiting gate operation within the surface code~\cite{gidney2019efficient}. To provide a range of estimates we consider the number of physical qubits and the resulting code distance required to implement the phase estimation algorithm on a surface code. We demonstrate assuming physical qubit error rates of 0.1\% that the ground state energy of the largest model Hamiltonian for the active space `X' of Cpd I can be assessed with approximately 4.6 million physical qubits in 73 hours of run time.  Anticipating improvements in qubit technology and reduction of error rates to 0.001\% the same computation preparing the ground state of Cpd I can be performed in approximately 500 thousand physical qubits and 25 hours of runtime.

The prevailing method for using a quantum computer to learn about the eigenspectrum of a molecular system is through phase estimation. The quantum computer simulates an operator related to the chemical system and learns spectral information without having to sacrifice accuracy beyond the basis set errors inherent to specifying the Hamiltonian. There are numerous variations of phase estimation that all come with different cost models depending on the type of Hamiltonian used and the type of physics being simulated. For chemical systems, recent work~\cite{berry2019qubitization, PhysRevResearch.3.033055, PRXQuantum.2.030305} has shown that applying phase estimation to the qubitization iterate~\cite{low2019hamiltonian} allows one to learn eigenvalues and prepare eigenstates with error no greater than $\epsilon$ by repeated application of the relevant quantum operator---the so-called qubitized quantum walk operator---
$\mathcal{O}(\lambda/\epsilon)$~\cite{PhysRevLett.121.010501} times where $\lambda$ is the $L_{1}$-norm of the coefficients of the Hamiltonian. The qubitized quantum walk approach seeks to perform phase estimation on an operator whose eigenvalues are $e^{\pm i \mathrm{arccos}(E)}$ where $E$ are eigenvalues of the original Hamiltonian~\cite{PRXQuantum.2.030305, PhysRevX.8.041015}. 

Three previous papers performed a full resource estimation on the number of physical qubits and run time requirements for simulating chemical systems in a molecular orbital basis using various forms of the qubitized quantum walk protocol~\cite{PRXQuantum.2.030305,PhysRevResearch.3.033055,berry2019qubitization}.  A detailed description of their differences, costs, and implementations can be found in Ref.~\cite{PhysRevX.8.041015} but here we highlight the differences relevant to quantum chemistry.  Most generically, for all qubitization schemes the walk operators can be implemented in Toffoli complexity scaling as $\mathcal{O}(\sqrt{\Gamma})$ and $\mathcal{O}(\sqrt{\Gamma})$ ancilla qubits where $\Gamma$ is the amount of information needed to specify a particular tensor factorization of the Hamiltonian coefficients where each tensor factorization can be implemented in $\mathcal{O}(N)$ Toffoli complexity. Using these costs, phase estimation can be implemented in a total gate complexity of $\tilde{\mathcal{O}}(\sqrt{\Gamma}\lambda/\epsilon)$\footnote{The $\tilde{\mathcal{O}}(.)$ indicates asymptotic scaling but dropping any parameters that scale as polynomials of logarithms in that parameter.}.  In all methods $\Gamma$ and $\lambda$ are non-trivially related to each other through the particulars of tensor factorization of the Hamiltonian coefficients. Thus, to find the best schemes for chemical systems we study the performance of three instances of qubitization schemes using different tensor factorizations in full detail to assess the overall scaling of each technique. 

The three Hamiltonian factorization schemes we compare are the single factorization (SF)~\cite{motta2021low, berry2019qubitization} (related to the Cholesky decomposition of the two-body electronic structure Hamiltonian~\cite{aquilante2011cholesky}), double factorization (DF)~\cite{motta2021low, PhysRevResearch.3.033055,peng2017highly}, and tensor hypercontraction (THC)~\cite{hohenstein2012tensor,hohenstein2012communication,lee2019systematically} with $\Gamma$ costs articulated in the Table~\ref{tab:qubitization_factorized_cost_table}. In both the DF and THC factorizations each tensor factor is evolved by rotating into a basis such that the central tensor is diagonal. This rotation costs $O(N)$ Givens rotations~\cite{PhysRevResearch.3.033055} which yield linear Toffoli complexity.  For THC the basis rotation is a projection into a larger basis with rank equal to the THC rank. In prior work utilizing the THC decomposition on the two electron integral tensor within the qubitization framework, a brute-force optimization scheme was used to determine the THC decomposition involving random restarts and direct gradient descent on the least-squares objective~\cite{PRXQuantum.2.030305}.  In this work we use a variation of prior THC decomposition workflows by starting with a symmetric canonical polyadic decomposition of the Cholesky vectors followed by $L_{1}$-regularized optimization of the THC factors. Full details of this protocol are described in Appendix~\ref{app:thc_cp3}. We validate that this new scheme reliably produces small $\lambda$ values for increasing system size. 
\begin{table}[H]
    \centering
    \begin{minipage}{0.85\textwidth}
        \caption{Tabulation of space and time complexity of the cost of performing phase estimation on qubitized quantum walk operators. Generically, qubitized quantum walks scale as $\tilde{\mathcal{O}}(\sqrt{\Gamma})$ in space and $\tilde{\mathcal{O}}(\lambda \sqrt{\Gamma}/\epsilon)$ where $\Gamma$ is the amount of information required to specify the Hamiltonian within a particular tensor factorization.  For the single factorization method (SF) $\Gamma = \tilde{\mathcal{O}}(N^{3})$. For double factorization $\Gamma = \tilde{\mathcal{O}}(N^{2}\Xi)$ where $\Xi$ is the average rank of the second factorization which is expected to scale as $O(N)$ in most regimes~\cite{motta2021low}.  For THC $\Gamma = \tilde{\mathcal{O}}(N^{2})$ assuming the THC rank grows linearly with system size.}
    \label{tab:qubitization_factorized_cost_table}
    \begin{ruledtabular}
    \begin{tabular}{ccc}
    Tensor Factorization Method   & Space Complexity (Logical qubits) &  Toffoli Complexity \\
    \hline
     SF  & $\tilde{\mathcal{O}}(N^{3/2})$    & $\tilde{\mathcal{O}}(N^{3/2}\lambda / \epsilon)$ \\
     DF & $\tilde{\mathcal{O}}(N\sqrt{\Xi})$ & $\tilde{\mathcal{O}}(N\sqrt{\Xi} \lambda / \epsilon)$ \\
     THC & $\tilde{\mathcal{O}}(N)$          &  $\tilde{\mathcal{O}}(N \lambda / \epsilon)$ \\
    \end{tabular}
    \end{ruledtabular}
    \end{minipage}
\end{table}

\subsection{Quantum resource scaling}

\begin{figure}[ht]
\centering
\includegraphics[width=0.35\textwidth]{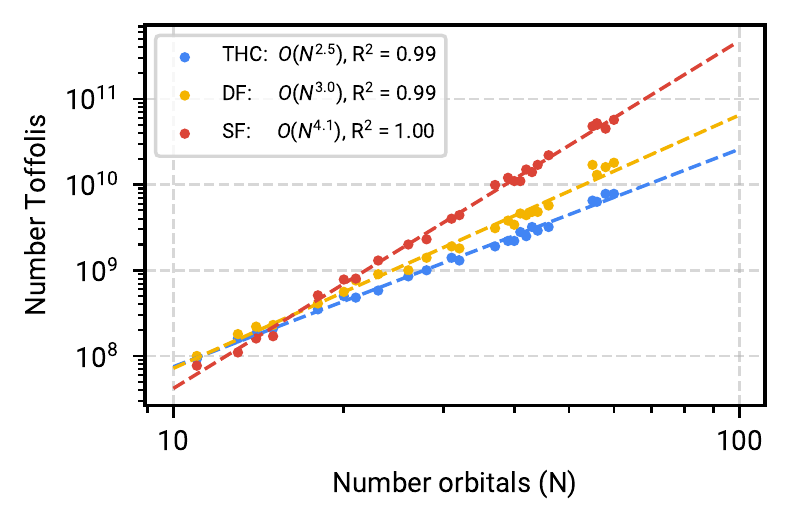}
\includegraphics[width=0.35\textwidth]{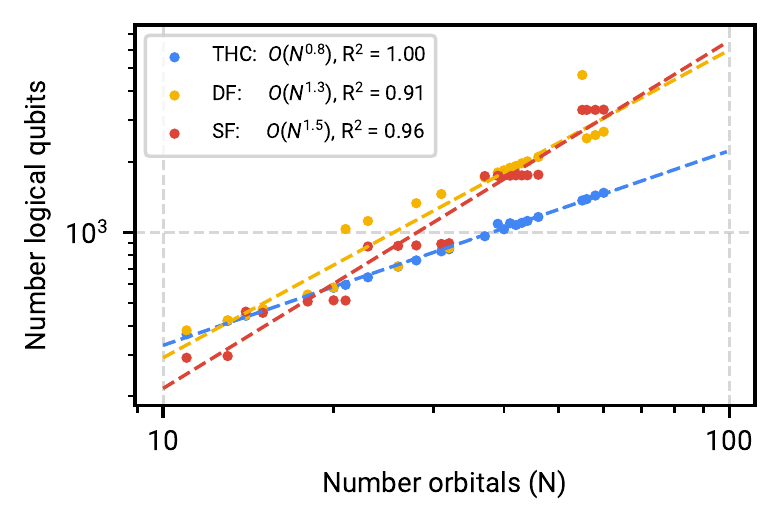}
\caption{ \textit{left}: Number of Toffolis as a function of the number of active space orbitals for the four heme compounds in this study, grouped by factorization algorithm: THC (blue), double factorization (DF; yellow), and single factorization (SF; red). Empirical scaling of the number of Toffolis as a function of orbitals $N$ is obtained by a least squares fit on the log-log plot.
\textit{right}: Number of logical qubits as a function of the number of active space orbitals for the four heme compounds in this study, grouped by factorization algorithm: THC (blue), double factorization (DF; yellow), and single factorization (SF; red). Empirical scaling of the number of qubits as a function of orbitals $N$ is obtained by a least squares fit on the log-log plot. \label{fig:toffoli_scaling_picture}
}
\end{figure}

Each factorization scheme requires a user-specified cutoff in terms of how accurately the two-electron integral tensor should be represented. This cutoff directly affects the scaling of the algorithms, and thus we use a heuristic to estimate the cutoff required for each Hamiltonian.  The heuristic we employ is using CCSD(T) with two-electron integrals reconstructed with a specific cutoff for the high spin geometries.  As validated previously, CCSD(T) is accurate in the active space for all compounds with a high-spin electron configuration. For example, to determine the sufficiently accurate THC rank, we perform THC decompositions with increasing rank until the CCSD(T) error is within one milliHartree (approximately 0.6 kcal/mol). \textcolor{black}{One milliHartree is selected for consistency with previous work~\cite{PRXQuantum.2.030305} but it may be necessary to select an even lower cutoff due to the fact that errors are additive between the total success probability of phase estimation, rounding in the procedure implementing the oracles associated with qubitization--QROM rounding--and truncation of two-electron integrals via approximate factorization.} From here, the $\lambda$ for each factorization is computed and subsequently input into the the detailed cost estimates for each step of the walk operator construction. Reference~\cite{PRXQuantum.2.030305} contains a full description of all costs associated with optimal implementation of each of the oracles in the qubitized quantum walk operator.  For reproducibility we provide a software tool that takes as input various tensor factorizations and computes the total resource requirements in terms of logical qubits and Toffolli counts.  The process of determining the THC rank cutoff for the largest active space of Cpd I is shown in Table~\ref{tab:cpd1_thc_convergence}. Considering the cutoff selected for all compounds the observed slope of the THC rank versus orbital number the is determined to be 4.7 (See Appendix~\ref{app:thc_cp3} Figure~\ref{fig:num_orbs_vs_thc_rank} for details).   
\begin{table}[H]
    \centering
    \caption{Large active space Cpd I using CAS((34$\alpha$, 29$\beta$), 58o) and showing the convergence of CCSD(T) for the high spin Hamiltonian as a function of the THC rank $M$. Highlighted in blue is the largest THC rank we consider for resource estimates.}
    \label{tab:cpd1_thc_convergence}
    \begin{ruledtabular}
        \begin{tabular}{c*{5}{D{.}{.}{2.1}}}
     $M$     &    \multicolumn{1}{c}{$||$ERI - THC$||$ }   &  \multicolumn{1}{c}{CCSD(T) error (mEh)}   &    \multicolumn{1}{c}{lambda}  &     \multicolumn{1}{c}{Toffoli count ($10^{9}$)} &      \multicolumn{1}{c}{logical qubits}    \\
     \hline
    140  &        4.7531   &          -107.80    &         260.4  &         4.3     &           1426   \\  
    160  &        2.4689   &           -28.33    &         326.7  &         5.6     &           1426   \\  
    180  &        2.3111   &           -12.62    &         331.0  &         5.8     &           1426   \\    
    200  &        1.5772   &           -5.27     &         352.3  &         6.3     &           1429   \\    
    220  &        0.46249   &           -0.84     &         385.0  &         7.0     &           1429   \\    
    240  &        0.70397   &           -1.45     &         377.6  &         7.0     &           1429   \\    
    260  &        0.52769   &           -0.20     &         382.8  &         7.3     &           1434   \\    
    280  &        0.45739   &           -1.29     &         384.8  &         7.4     &           1434   \\    
    300  &        0.36635   &           -1.17     &         387.4  &         7.6     &           1434   \\  
  \color{blue} 320  &     \dcolcolor{blue} 0.31506  &     \dcolcolor{blue}     0.10     &     \dcolcolor{blue}    388.9  &    \dcolcolor{blue}   7.8     &      \dcolcolor{blue}    1434   \\    
    340  &        0.24618   &           -0.14     &         390.9  &         8.0     &           1434   \\    
    360  &        0.22478   &           -0.71     &         391.6  &         8.2     &           2156   \\    
    380  &        0.19167   &           -0.83     &         392.5  &         8.3     &           2158   \\    
    400  &        0.15139   &           -0.45     &         393.7  &         8.5     &           2158   \\    
    \end{tabular}
  \end{ruledtabular}
\end{table}
As previously described, the true scaling of qubitization for each factorization depends largely on how well the various factorizations can use the existing structure in the two-electron integral tensors and compress the information.  Using the progressively larger active spaces and resource estimates in terms of Toffolli gate and logical qubit counts for each factorization we can assess the scaling and extrapolate to larger size factorizations not studied in this work.  The extrapolations for number of Toffoli gates and number of logical qubits required are shown in Figure~\ref{fig:toffoli_scaling_picture}.

\subsection{Compilation into Surface Code}

\newcommand{\chemftrsrc}[1]{\href{https://github.com/ncrubin/chemftr/blob/main/src/chemftr/{{#1}}}{ \path{chemftr/#1}}}

In order to perform the required number of gates for phase estimation the quantum state must be protected against errors through a quantum error correction protocol.  The surface code is one such protocol that can be implemented on a two-dimensional array of qubits and requires physical error rates no worse than a 0.5\% error rate~\cite{PhysRevA.86.032324} .  In the surface-code one can make trade-offs between space and time, i.e. changing the length of time required for the computation at the expense of using more physical qubits. Significant controlling factors for this are the number of Toffoli factories and how these resource factories are implemented. 

We start with analyzing the runtime requirements for simulating the largest 'X' Hamiltonians. For these systems dynamic correlation corrections may be small and are thus more likely to result in accurate spin-gaps. To estimate the cost of executing phase estimation in the surface code, we start from the number of data qubits and Toffoli gates required, determined from the previous section based on a THC factorization of the two-electron integrals. We assume that four magic state factories are being used and the execution time is determined by the Toffoli count, due to being bottlenecked waiting for magic state factories. Higher factory counts would require a much different analysis on routing overheads. We assume space usage is determined by the number of logical data qubits, the number of magic state factories, and a 50\% overhead for routing. Finally, we assume a physical per-gate error rate of 0.1\%, a surface code cycle time of 1 microsecond, and a control system reaction time of 10 microseconds.

These assumptions are used to determine a variety of different configurations (code distances and factory layouts) and selects the configuration that uses the least spacetime volume while ensuring that the quantum computation corrects all errors at least 90\% of the time. The code to do this estimation is provided as a submodule of OpenFermion \cite{mcclean2020openfermion}.  For physical qubit error rates below 0.1\% we use AutoCCZ factories from Ref.~\cite{gidney2019efficient} and T factories from Ref.~\cite{fowler2018low}. It estimates the failure rate of the factories, and the failure rate of logical data qubits, in order to get an overall failure rate for the entire algorithm for a given configuration.

\begin{figure}[H]
    \centering
    \includegraphics{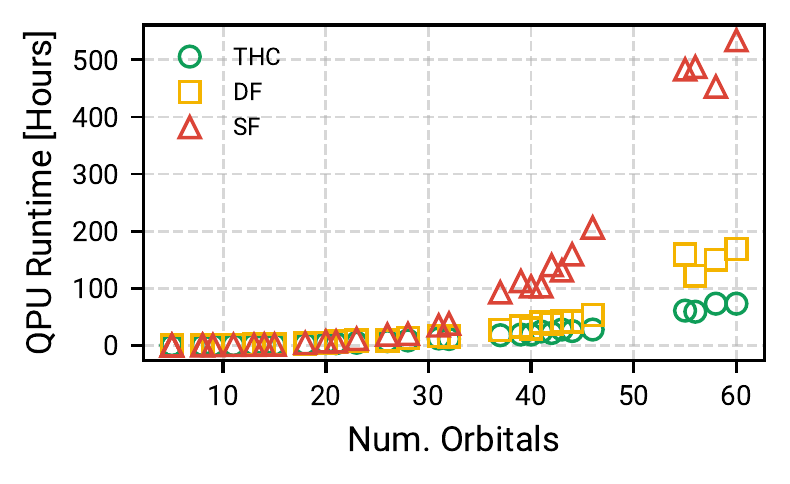}
    \caption{Phase estimation runtimes with physical qubit error rate of 0.1\% and four Toffoli factories while assuming a 1 $\mu$s surface code cycle.}
    \label{fig:run_times_all_systems}
\end{figure}

\begin{figure}[H]
    \centering
    \includegraphics[width=0.5\textwidth]{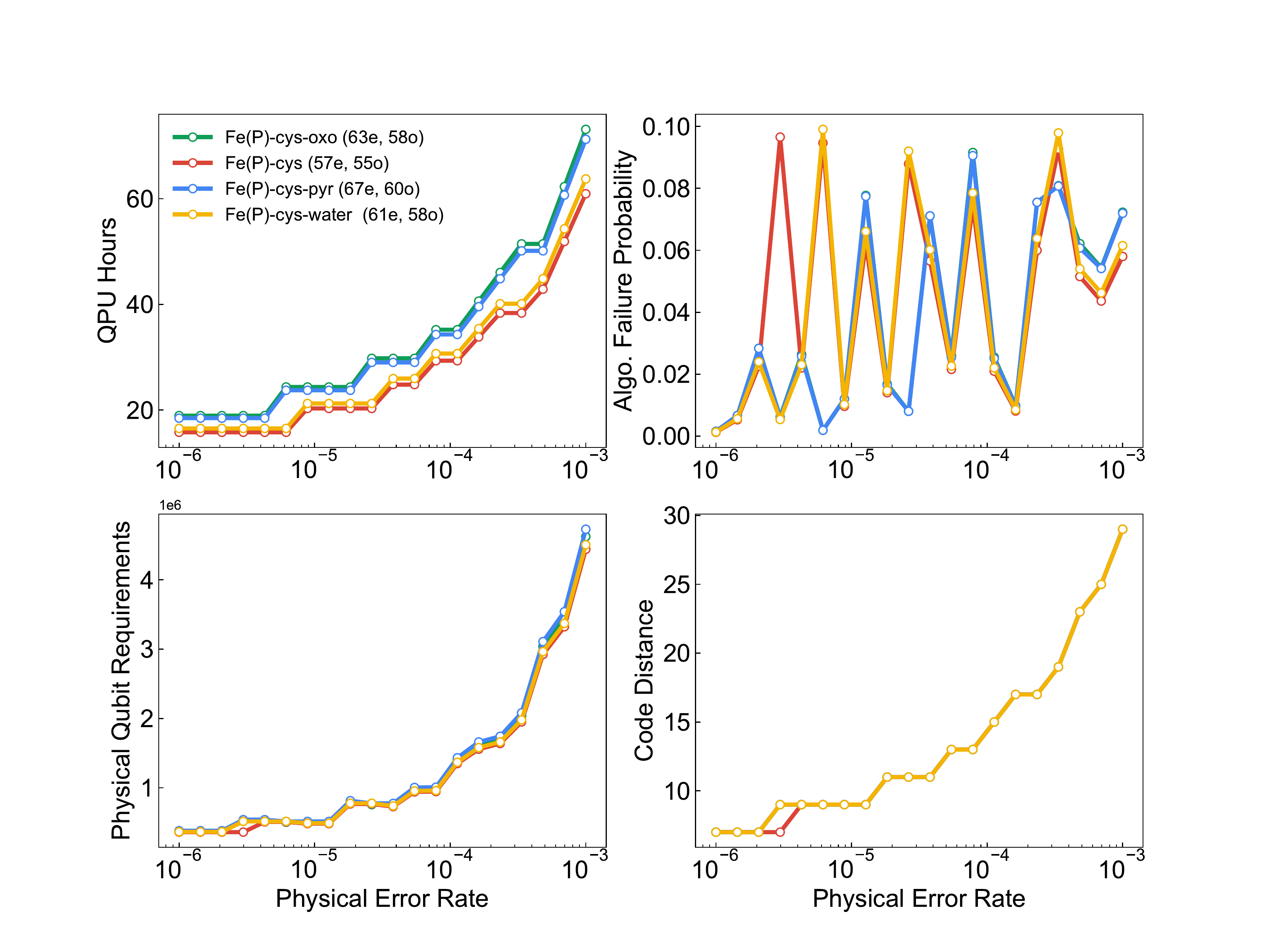}
    \caption{Comparison of compiled resource requirements as a function of physical error rate: \textit{upper left}: runtime, \textit{upper right}: algorithm failure probability, \textit{lower left}: physical qubit requirements, \textit{lower right}: required code distance. We note that error rates on the order of $1.0\times 10^{-6}$ are highly unrealistic and if achieved would precipitate using a different error correction protocol.}
    \label{fig:digital_craig}
\end{figure}

For a physical error rate of 0.1\%, four factories, and the aforementioned surface-code timings the optimal configuration was AutoCCZ magic state factories with a level-1 code distance of 19 and level-2 code distance of 31, and logical data qubits with a code distance of 29. This requires 4,624,440 physical qubits and a total runtime of 73 hours.  For these estimates we assume that the space-time requirements (qubits and Toffolis) are translated directly into error corrected requirements which is an over estimation of the needed resources. Thus these estimates should be viewed as upper bounds assuming other timing criteria are met by real machines.  In Figure~\ref{fig:run_times_all_systems} we plot the runtime scaling for phase estimation performed on all active space calculations with different tensor factorizations. The data shows that despite higher prefactors for tensor hypercontraction qubitization the asymptotic scaling advantage appears even at small sizes.
As a forecasting exercise we can make runtime, physical qubit, and code distance estimates as a function of physical qubit error rate.  This provides insight into the scenario of qubit error rates becoming substantially better.  We again make the assumption that surface code cycle and reaction times are 1~$\mu$s and 10~$\mu$s and we use four Toffoli factories.  In Figure~\ref{fig:digital_craig} we show the decreased runtime, physical qubit requirements, and code distances as a function of physical error rate.  In the event we have extremely low gate error rates of 0.001\% solving the largest system will take only 500 thousand qubits and 25 hours of runtime.
\subsection{Demarcating the quantum advantage boundary}
\begin{figure}[H]
    \centering
    \includegraphics[width=0.30\textwidth]{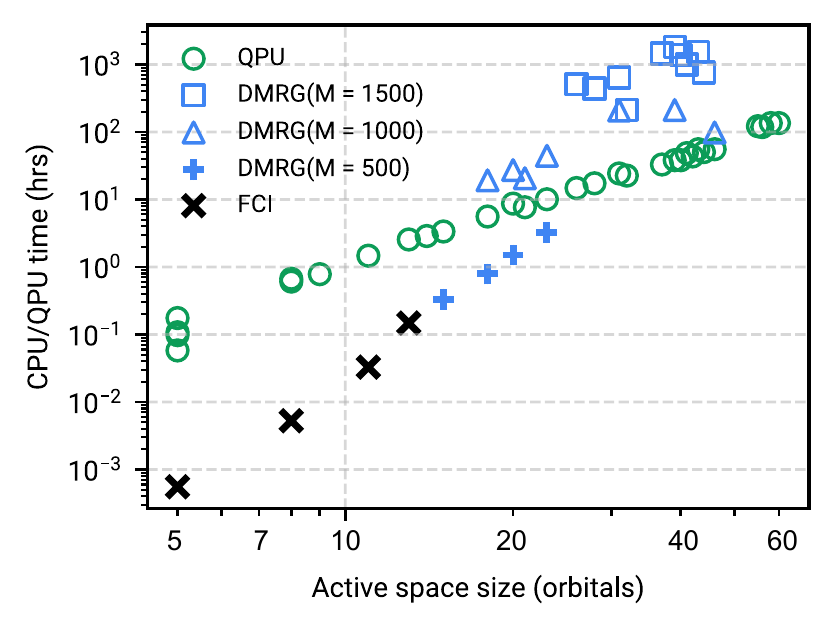}
    \includegraphics[width=0.30\textwidth]{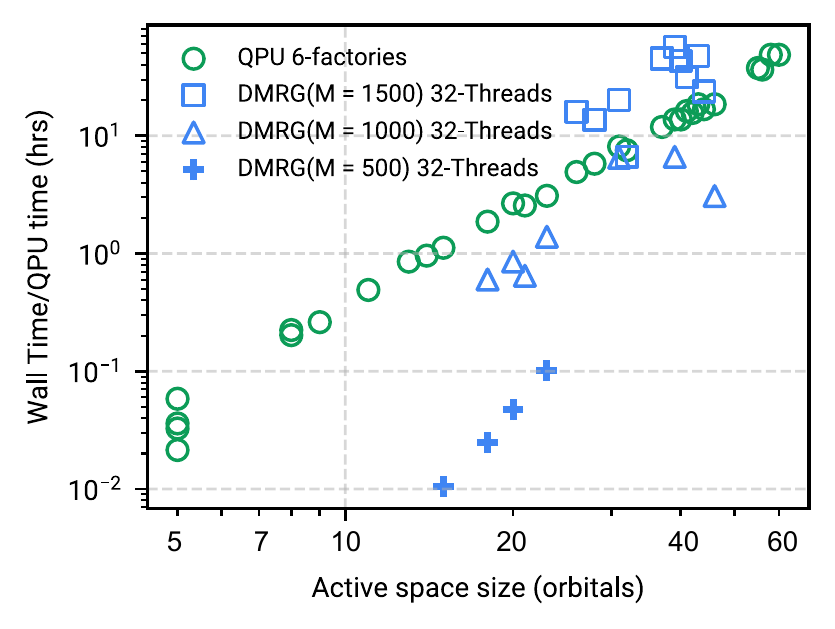}
    \includegraphics[width=0.30\textwidth]{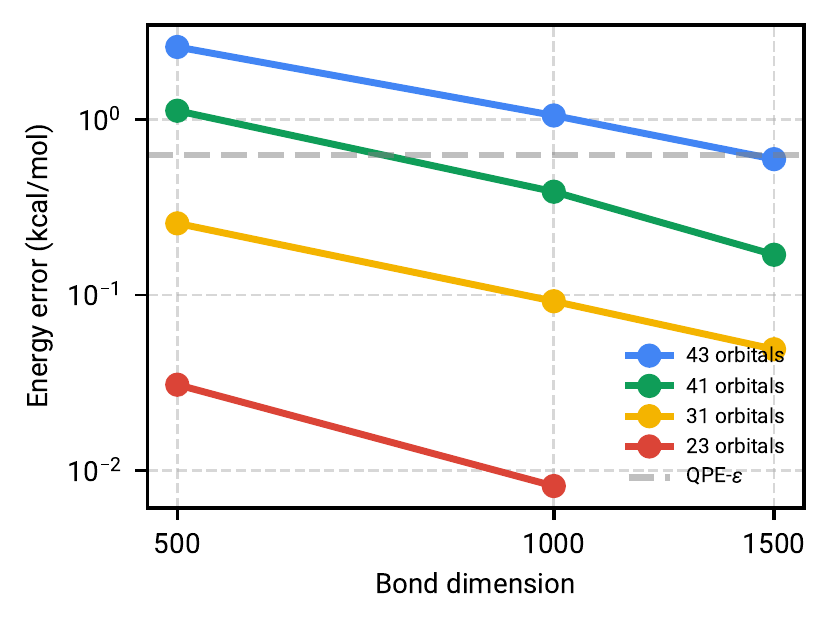}
        \caption{\textit{left:} CPU time for CASCI/DMRG calculations on different active spaces and QPU time to perform phase estimation on the THC-decomposed active-space Hamiltonian. Times are determined by wall clock time multiplied by the number of threads.  QPU time is determined assuming a 0.1\% gate error rate and two Toffoli factory and neglects repetitions needed due to potentially small initial state overlap.  Both methods, save CASCI, are parallelizable to some extent by using more resources. \textit{center:} Same timings as \textit{right} but now considering 32 threads for DMRG and 6 Toffoli factories for the QPU. \textit{right:} DMRG energy at fixed bond dimension relative to extrapolated energy for various number of orbitals.}
    \label{fig:qvsc_timings}
\end{figure}

Using timings from DMRG calculations and estimated runtimes on an error corrected quantum computer we can compare the corresponding CPU and QPU time requirements as a function of active space size to investigate the potential for a simulation advantage.  In Figure~\ref{fig:qvsc_timings} we plot the measured timings of DMRG calculations on all active spaces for different values of bond dimension ($M$) and the estimated runtimes for the quantum computing to perform phase estimation. The QPU runtimes are estimated using two Toffoli factory which corresponds to 4.9 million qubits and 135 hours of run time for the largest systems.  DMRG runtimes are computed as the wall clock time multiplied by the number of threads. In practice the run time of either calculation can be reduced by using more cores or more Toffoli factories, respectively. To put the DMRG timings in context relative to QPU timings we need a notion of how accurate they are at fixed bond dimension.  Due to the lack of convergence in any of the DMRG or coupled-cluster calculations for the spin-gap we instead compare to the extrapolated DMRG energy for a given active space. This comparison makes the assumption that the extrapolated energy at any bond dimension is highly accurate. Figure~\ref{fig:qvsc_timings} suggests that for large bond (M$\geq$1000) dimension and large active space  quantum phase estimation already has a computational advantage when the task is simulating the ground state energy.  

While DMRG provides access to reduced density matrices (and therefore other observables), phase estimation only provides estimates of the energy and further processing is required to compute other quantities. This is important for energetic quantities as well, because corrections for dynamic correlation outside of the active space usually require reduced density matrices (1-, 2-, 3-, and sometimes 4-particle). These corrections, such as the NEVPT2 used in this work, can allow for smaller active space calculations in principle. This difference highlights a need for quantum algorithms addressing dynamic correlation.  Extrapolating out the number of qubits and Toffoli gates required for CYP active space models at 500 orbitals (which is approximately the number of orbitals used for the full space coupled cluster calculations) we would require approximately 9000 logical qubits and 1.5 trillion Toffoli gates to perform phase estimation on the entire space.  Thus, the development of quantum algorithms for addressing dynamic correlation is an important step towards chemical computational relevancy.  

\begin{figure}
    \centering
    \includegraphics{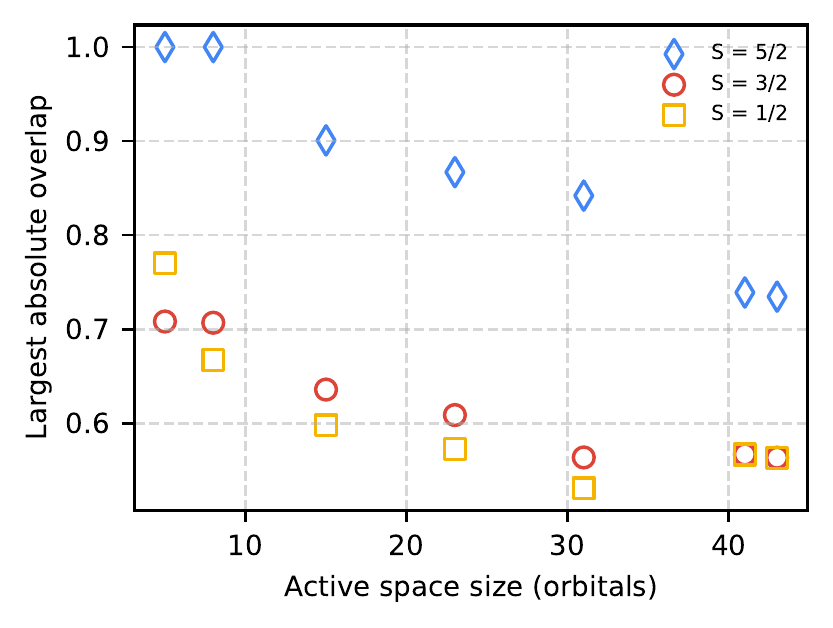}
    \caption{Largest computational basis state overlap with the ground state DMRG $M=1500$ wavefunction.}
    \label{fig:overalps}
\end{figure}

The success probability for phase estimation also relies on overlap $S$ of the initial state with an eigenstate of the Hamiltonian.  Therefore, a full timing comparison would ideally factor the overlap dependency as a $poly(S^{-1})$ multiplying prefactor which can be improved with knowledge of the gap~\cite{berry2018improved}.  To have an idea of the size of this prefactor for Cpd I we have computed the determinant used in the exact diagonalization basis, also known as a computational basis state, that has the largest overlap with the DMRG $M=1500$ bond dimension wavefunction for systems $A$ through $G$.  In Figure~\ref{fig:overalps} we plot the overlap as a function of active space size for all three spin states considered.  As expected, the largest overlap decays by a factor of two going from a system size of 10 qubits to 84 qubits but never falls to a value that would be problematic for phase estimation.

\section{Numerical support for a multireference treatment of p450}\label{sec:justifications}
The electronic structure of CYPs has been extensively studied computationally in the literature using both polynomial-scaling methods such as DFT \cite{Shaik:2010} and using active space methods \cite{Schoneboom2005,Altun2008,Chen2011}. Although DFT-based simulations have successfully revealed many aspects of potential reaction mechanisms \cite{Shaik:2010}, it has been argued that there is a need for application of active space based methods to achieve high accuracy for spectroscopic properties \cite{Schoneboom2005} and spin-state orderings \cite{Altun2008, Chen2010, Chen2011}. Our cost assessment for simulation on classical and quantum hardware in the previous sections has been based on this assumption. In the following, we will provide further numerical evidence that predictive simulation of the catalytic cycle of CYPs requires a multireference treatment.
We investigate the multiconfigurational character of the model compounds with full-space (no active space) calculations by considering four commonly-used diagnostics: single-determinant spin contamination, $\max (|t_1|)$ and $\max (|t_2|)$ diagnostics from CCSD \cite{cramer2006theoretical,liakos2011interplay,Jiang2012-qx,radon2014spin,margraf2017single,cheng2017bond,feldt2019limits}, and deviations from ideal spin multiplicity as given by natural orbital occupation numbers from the $\kappa$-OOMP2 method.  For completeness we provide a detailed discussion on the chemical meaning of each metric in Appendix~\ref{app:muliref_stats}.

\begin{figure}[ht]

\centering
\includegraphics[width=0.95\textwidth]{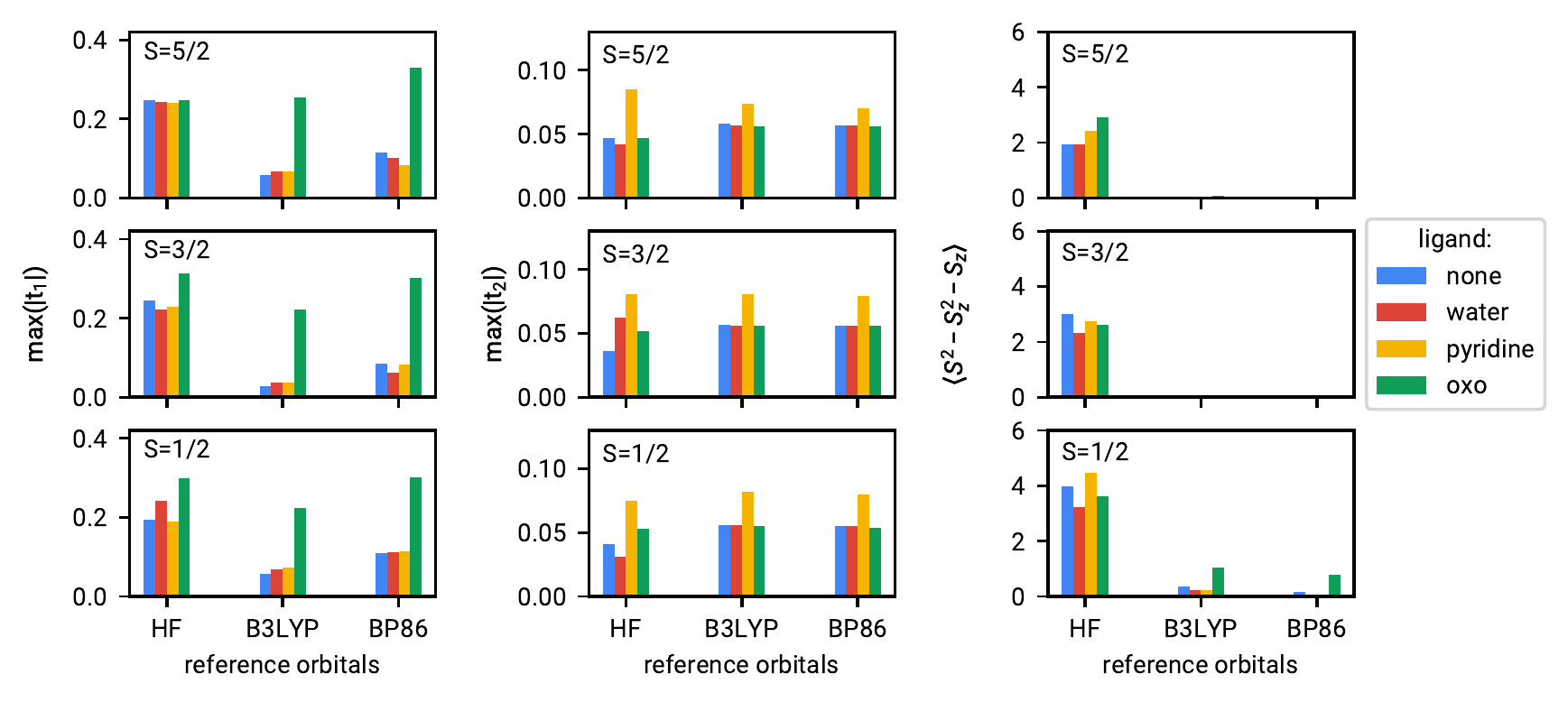}
\caption{\textit{left}: Impact of reference orbital on the $\max \left(|\mathrm{t}_1|\right)$ values from CCSD. \textit{center}: Impact of reference orbital on the $\max \left(|\mathrm{t}_2|\right)$ values from CCSD. \textit{right}: Spin contamination, $\langle S^2 - S_z^2 - S_z \rangle$, as a function of reference orbitals. For most species, use of KS reference orbitals removes nearly all of the spin contamination in the SCF, in contrast to the highly-spin contaminated HF orbitals. The doublet (S=1/2) and quartet (S=3/2) states for the oxo ligand in Cpd I remain somewhat spin contaminated despite using KS reference orbitals. }
\label{fig:mr_diagnostics}
\end{figure}

First we examine the $\max (|t_1|)$ diagnostic from CCSD which characterizes the \emph{quality} of reference orbitals \cite{cramer2006theoretical,liakos2011interplay,Jiang2012-qx,cheng2017bond}. This quantity is plotted in Figure~\ref{fig:mr_diagnostics} for orbitals obtained from two flavors of Kohn-Sham density functional theory and Hartree-Fock theory.
With Hartree-Fock orbitals, all compounds have a relatively large $\max(|t_1|) > 0.19$, suggesting that Hartree-Fock is a poor reference for these systems. However, using Kohn-Sham orbitals, which include dynamic correlation effects, is sufficient to yield a significantly lower value of $\max(|t_1|)$ for all compounds except for the iron-oxo containing Cpd I. This shows that for three of the four compounds (empty pentacoordinate, resting, and pyridine inhibited) the relatively large $\max(|t_1|)$ can be corrected by including some dynamic correlation in the reference orbitals. In other words, with an improved SCF reference, the systems appear to be single-reference. However, this is not the case for Cpd I given the KS orbitals considered here, which suggests that a single-reference approach may struggle to capture the true nature of electron correlation.

Another commonly used diagnostic of multiconfigurational character is spin-symmetry breaking in the wave function \cite{Jiang2012-qx,radon2014spin}, seen in Figure~\ref{fig:mr_diagnostics}. The reference orbitals given here are from unrestricted calculations, and may deviate from ideal eigenstates of $S^2$, shown by the nonzero expectation value of spin contamination, $\langle S^2 - S_z^2 - S_z \rangle$. Like large values of $\max (|t_1|)$, spin contamination is often used to identify underlying multiconfigurational character. For example, it has been used to support the claim the C$_{60}$ is polyradical \cite{jimenez2014polyradical}, despite experimental evidence to the contrary. While spin contamination can, in some cases, be a genuine marker of multiconfigurational character (so-called ``essential’’ symmetry breaking), it may also be the result of ``artificial’’ symmetry breaking, due to the lack of dynamic correlation. We observe that for Hartree-Fock orbitals, the spin contamination is large for all compounds and all spin states investigated. However, upon switching the reference orbitals to Kohn-Sham orbitals, we note that the purported spin contamination is essentially eliminated, barring some spin contamination in the doublet of Cpd I. This indicates, once again, that by carefully choosing an appropriate SCF reference (here with KS orbitals) it is possible to properly describe the system with single reference methods. In most cases, DFT greatly reduces large $\max (|t_1|)$ values as well as eliminates spin contamination, showing the spin contamination observed in the HF references as ``artificial’’. The exception is the doublet state of Cpd I.

A more rigorously justified diagnostic of strong correlation (or multiconfigurational character) is the $\max (|t_2|)$ diagnostic from CCSD \cite{cramer2006theoretical,liakos2011interplay,Jiang2012-qx,radon2014spin,margraf2017single,cheng2017bond,feldt2019limits}. Large amplitudes from the double excitations, generally considered when $\max (|t_2|) > 0.05-0.1$, are relatively insensitive to the underlying reference orbitals. Unlike single excitations in CCSD, which are essentially orbital rotations, doubles amplitudes directly capture the lowest-order pairwise correlations between electrons. We observe that for all compounds and all spin states, the value of $\max (|t_2|)$ is more-or-less constant, roughly centered around $\max (|t_2|) \approx 0.05$. Depending on the threshold for $\max (|t_2|)$ (we would consider the more stringent and common $\max (|t_2|) > 0.1$ to be a marker of strong correlation), the data show that none of the various model compounds are particularly strongly correlated. In light of the other diagnostics and the fact that using KS orbitals appears to restore most compounds to a reasonable single-reference configuration, we see no evidence that the P450 model compounds are strongly correlated. 

\begin{figure}[ht]

\centering
\includegraphics[width=0.4\textwidth]{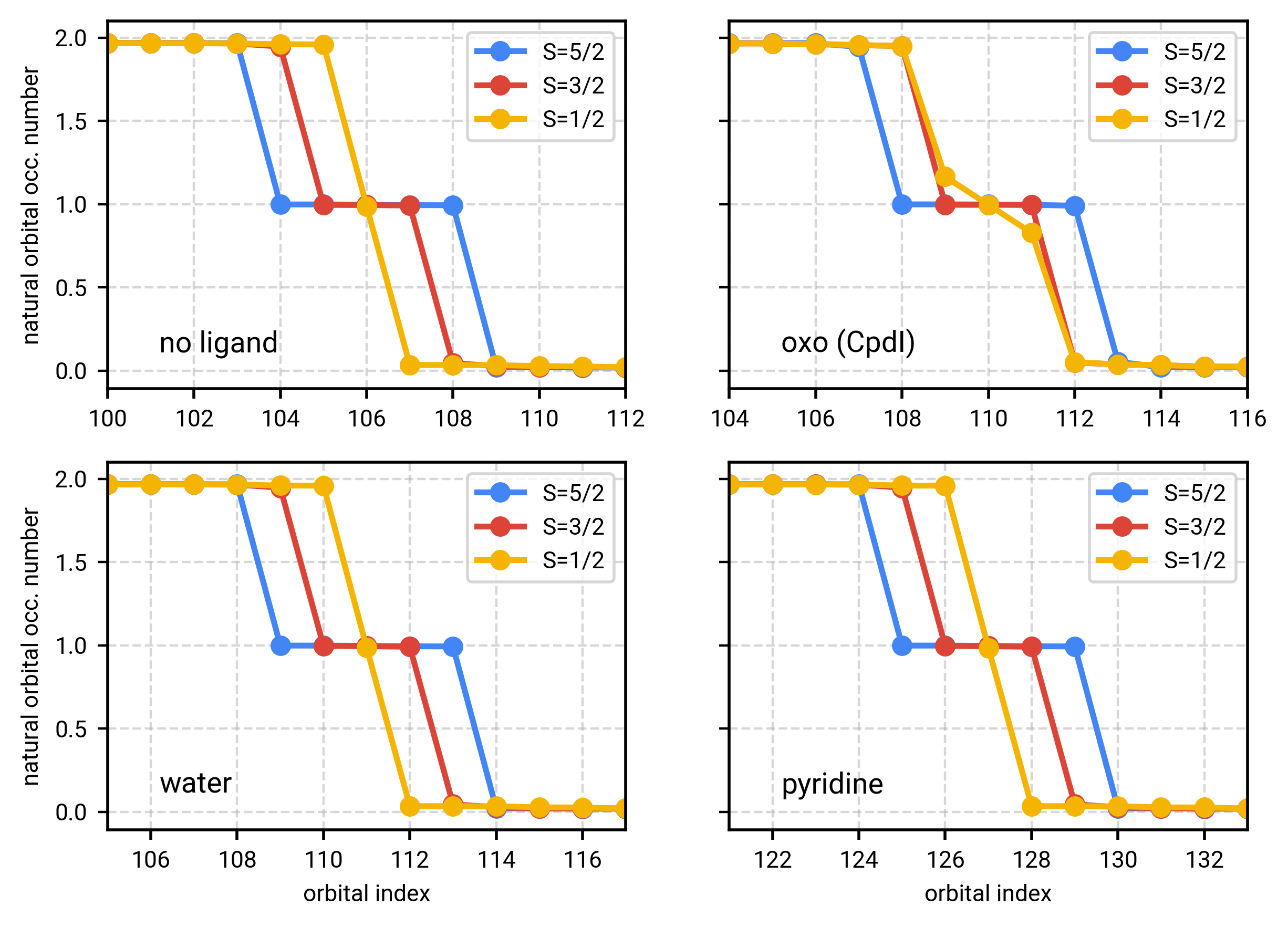}
\caption{$\kappa$-OOMP2 ($\kappa = 1.1$) natural orbital occupation numbers for the four compounds and three associated spin states. All calculations are performed at the high-spin (S=5/2) geometry obtained from a B3LYP geometry optimization, as detailed in the text. The open-shell character of all compounds corresponds to the idealized spin multiplicity, with the lone exception being the iron-oxo Cpd I doublet, which shows three open shells for the doublet state, in agreement with the DMRG calculations in this work as well as previous calculations indicating triradical character, such as in Ref.~\cite{Chen2011}.}
\label{fig:oomp2}
\end{figure}

Finally, we examine the natural orbital occupation numbers (NOONs) from  the correlated 1PDMs of $\kappa$-OOMP2 calculations as shown in Figure~\ref{fig:oomp2}. For the three compounds that we expect to have largely single-reference character, the sextet, quartet, and doublet states have 5, 3, and 1 open-shells respectively as expected. However, the ``doublet" state of Cpd I has 3 NOONs close to one indicating a state of triradical character which is consistent with prior calculations~\cite{Chen2011}. Though the $\max (|t_2|)$ diagnostic does not indicate that the Cpd I doublet is any more strongly correlated than the other model compounds, such a state cannot be faithfully represented by a single determinant. Though one can compute some properties, such as the nuclear gradient, with symmetry-broken states, it is often difficult to accurately compute spin-dependent properties, like the spin gap, without spin-pure states.

Taken together, we, like others, conclude that P450 appears to have a multiconfigurational electronic structure. A treatment of electron correlation that can approach the exact solution is essential to resolve the nearly degenerate quartet and doublet states. In other words, to achieve the high accuracy needed to assess spin energetics and other properties, multireference methods are required. Doing this reliably is not an easy task. Though Cpd I is not ``strongly correlated" in the sense that it does not have a large number of strongly entangled orbitals, there is no easy way to do reliable simulation, and the need for multireference methods transcends the set of strongly correlated problems. This motivates the need to develop quantum algorithms and methods that can scale to sizes large enough to balance dynamic and static electron correlation.

\section{Conclusion}
In order to demarcate the quantum advantage boundary for biologically relevant compounds we have performed a detailed characterization of the classical and quantum resources required to accurately describe the electronic structure of the P450 enzyme active site. This work highlights the demand of careful application of a variety of classical chemical methods and that even in regimes where a problem is not ``strongly correlated'' quantum computers can potentially provide advantage due to the need to treat dynamic and static correlation in a balanced way.  We also determine that circumventing the traditional strategy of partitioning dynamic and static correlation by using a quantum computer to simulate the entire problem is unlikely to be feasible due to the very large number of resource states needed to execute phase estimation. These findings suggest that further development of fault tolerant algorithms, either in observable extraction or scaling, are necessary for quantum computers to be transformative for simulating electronic structure where modest multireference character is augmented by a large dynamic amounts of correlation. 

To make the aforementioned conclusions we analyzed the multireference character through a variety of classical electronic structure methods which help to clearly determine the computational frontier of CYP simulation.  The necessity of 
using multireference methods is supported by examining spin-contamination, defined by $\langle S^2 - S_z^2 - S_z \rangle$, as well as three other metrics from correlated wave function theory, namely, max($|t_{1}|$) and max($|t_{2}|$) from CCSD, as well as natural orbital occupation numbers from regularized $\kappa$-OOMP2. From these metrics, we found that Cpd I displayed some multireference character. DMRG calculations confirm the triradical character involving three open-shells in the Cpd I doublet corresponding to Fe-O $\pi^{*}_{xz/yz}$ orbitals and a lone pair orbital on the sulfur atom of the Me-S ligand emulating the full coordinating cysteine ligand. This triradical character is consistent with previous multireference calculations on Cpd I\cite{Altun2008,Chen2010,Chen2011}.  Despite a clear characterization of non-trivial open-shell electronic structure in Cpd I, the ground-state spin state and spin gaps remain elusive due to the need to treat dynamic and static correlation on equal footing.  

Analysis of the quantum resources required to simulate these systems indicated that of the three Hamiltonian factorizations (SF, DF, and THC factorization) used in qubitized phase estimation, tensor hypercontraction consistently outperformed the other two. Compilation into the surface-code provided an upper bound runtime estimate for executing phase estimation.  Most notably, under realistic hardware configurations we predict that the largest models of CYP can be simulated with under 100 hours of quantum computer time.  A direct runtime comparison of qubitized phase estimation shows a more favorable scaling than DMRG, in terms of bond dimension, and indicates future devices can potentially outperform classical machines when computing ground state energies. Extrapolating the observed resource estimates to the full Cpd I system and compiling to the surface-code indicates that a direct simulation of the entire system could require 1.5 trillion Toffoli gates--an unfeasible number of Toffoli gates to perform. 

The classical benchmarking of CYP compounds demonstrates the need to account for dynamic correlation and the quantum cost estimates detailing the requirements for a high accuracy simulation encourages further development of quantum algorithms to address multireference quantum chemistry beyond the strong correlation regime.  Furthermore, this work demonstrates that classical calculations play an important role in guiding quantum algorithms research and are essential for defining the computational frontier for chemistry.

\section*{Acknowledgements}
The authors thank Leon Freitag for code contributions and discussions on active space construction, and Clemens Utschig-Utschig for discussions regarding the content and vision of the paper.

\section*{Data Availability}
For reproducibility, we share data and code used in this work on a public Zenodo repository (10.5281/zenodo.5941130), including the molecular geometries and active space Hamiltonians, along with the inputs to reproduce the calculations. All software used for Hamiltonian tensor factorizations and phase estimation resource estimates can be found in the \texttt{resource\_estimates} module of OpenFermion~\cite{mcclean2020openfermion} commit number \\ \texttt{cf53c063d0f124a02ff8776bb7f8afb110d4bde6}. To perform the tensor hypercontraction factorization with the code in OpenFermion an interface,\texttt{pybtas}, to BTAS~\cite{BTAS} commit \texttt{5702259d5d207fe5a8e0c975c3cf1f610dcf381a} is required.  The pybtas library is included in the Zenodo repository and can be found at \url{https://github.com/ncrubin/pybtas}. The THC factors obtained from this work are also available in the Zenodo repository.
\bibliographystyle{apsrev4-2}
\bibliography{p450,AddedRefs}

\appendix
\section{Electron correlation and the computational cost of electronic structure calculations}\label{app:muliref_stats} 

The nature of the electronic correlation in a particular system influences the choice of appropriate electronic structure method. For systems with wave functions that are dominated by a single electronic configuration—that is, single reference—it is generally sufficient to treat the electronic correlation using low-order perturbation theory, for example, with the hierarchy of coupled cluster methods. These methods quantitatively correct for the lack of electron correlation by incorporating excited Slater determinants into the wave function. These excited Slater determinants generally yield small contributions, in the sense that that their wave function amplitudes are relatively small compared to the dominant zeroth-order reference. That said, the number of excited determinants that need to be considered may be large. When a large number of small corrections to the wave function is required to treat electron correlation, the correlation is often referred to as ``dynamic correlation.'' Dynamic correlation is straightforward (though not necessarily cheap!) to systematically improve, simply by increasing the order of theory.

In contrast, for systems where no single dominant configuration is sufficient to describe the wave function even qualitatively, low orders of perturbation theory will fail. Several electronic determinants are required, and their contribution to the overall wave function will be of roughly equal magnitude. Such a system is considered ``multireference'' or ``strongly correlated,'' and the electron correlation is often referred to as ``static correlation'' ``Static'' and ``dynamic'' correlation can be viewed as two classification extremes that are useful for developing methods.  For many systems, careful consideration of both types of electronic correlations are required, and demand methods that can balance accuracy of describing ``static'' and ``dynamic'' correlation. As both are required to treat chemical systems with sufficient accuracy, it is important to not treat one type of correlation more accurately than the other: insufficient accuracy in one will lead to errors computed properties, regardless of how accurate the other correlations were computed. A balanced treatment of electron correlation is not a trivial task and imbalances may lead to difficulties in describing the spin-state ordering, as we will show in the model systems discussed in this work.

Generally, it is not possible to know \emph{a priori} whether or not a chemical system is either single or multireference. Though several heuristics have been established (e.g. does the system contain multiple metal sites?), the only reliable way to determine if a problem is multireference is to subject it to calculation. This presents a quandary: on the one hand, one wants to avoid expensive multireference methods where possible, yet on the other hand, the most reliable way to determine if such methods are necessary is to employ said methods. This, coupled with the fact that there is no clear single ``measure'' of strong correlation, makes scoping the problem very difficult. To this end, several probes of strong correlation have been proposed that, when taken together, give one reasonable confidence that a multireference approach is or is not necessary. 

In this work, we examine the merits of four commonly used probes, or diagnostics, of strong correlation: 
\begin{enumerate}
    \item{} Spin symmetry breaking. It is generally thought that spin-symmetry broken wave functions indicate the presence of strong correlation, for example, beyond the Coulson-Fisher point in homolytic bond dissociation. This can be quantified by computing the spin-contamination, defined as the $\langle S^2 - S_z^2 - S_z\rangle$ expectation value. However, the use of spin-contamination as a diagnostic of strong correlation is not as clear cut. Recent work~\cite{lee2019distinguishing} has shown that spin-symmetry broken mean-field solutions may be ``cleaned up'' by either by switching to a different SCF reference (such as one based on Kohn-Sham DFT) or by utilizing an orbital-optimized method in the presence of dynamic correlation, such as $\kappa$-OOMP2 \cite{Lee2018}. References that can have their spin symmetry ``restored'' in this manner are referred to as ``artifically'' spin-symmetry broken solutions. Those that cannot, on the other hand, show ``essential'' spin-symmetry breaking and do indeed indicate some measure of multireference character or strong correlation. As an example, prior work on the bare iron porphyrin complex details extensive calculation providing evidence that the complex is single reference~\cite{lee2020utilizing} and arguing that spin-contamination is largely a product of ``artificial'' symmetry breaking from not including an accurate treatment of dynamic correlation at the single-determinant level of theory.
    \item{} Measures based on the coupled cluster singles amplitudes. While norms of the CCSD $t_1$ amplitudes, such as the T$_1$ or D$_1$ diagnostic \cite{lee1989diagnostic,leininger2000new,lee2003comparison}, are commonly used to evaluate whether a system is strongly correlated, these measures are more of a measure of the quality of the underlying orbitals. The reason is that in coupled cluster theory, the single excitations act as an orbital rotation operator. In many cases, choosing a more suitable reference is sufficient to eliminate large $t_1$ norms. One additional problem with relying on norms of the $t_1$ amplitudes is that it masks heterogeneity in the singles amplitudes, i.e. if only a small number of singles amplitudes are large, the norm will not be representative. For this reason, we report the $\max (|t_1|)$ value, as has been done elsewhere \cite{cramer2006theoretical,liakos2011interplay,Jiang2012-qx,cheng2017bond}.
    \item{} Measures based on the coupled cluster doubles amplitudes. In contrast to the singles, measures based on the doubles amplitudes have a clear connection to electron correlation, as they generate configurations that cannot be obtained by merely rotating the reference wave function. Moreover, doubles amplitudes only have a weak dependence on the underlying reference. Here report the $\max (|t_2|)$ value \cite{cramer2006theoretical,liakos2011interplay,Jiang2012-qx,radon2014spin,margraf2017single,cheng2017bond,feldt2019limits}. When this value is large (e.g. $ \max (|t_2|) > 0.1$) this is a reliable indication the system is multireference. (Recall that due to intermediate normalization in coupled cluster theory, the reference amplitude is always unity.)
     \item{} Natural orbital occupation numbers from correlated methods. Natural orbital occupation numbers (NOONs), which are the eigenvalues of the 1PDM, can deviate from the idealized open-shell character when using correlated 1PDMs. In this case, the NOONs indicate that additional open shells are present (e.g. polyradicaloid character) which indicates unambiguously that the chemical system is multireference. In this work, we consider NOONs from the $\kappa$-OOMP2 method, which provides an affordable route to evaluating multireference character.
\end{enumerate}
Although this list is by no means exhaustive, it does present some commonly utilized diagnostics used by the electronic structure community to evaluate multireference character.

\section{Cytochrome P450}\label{app:P450} 
\begin{figure}[ht]
\centering
\includegraphics[width=0.75\textwidth]{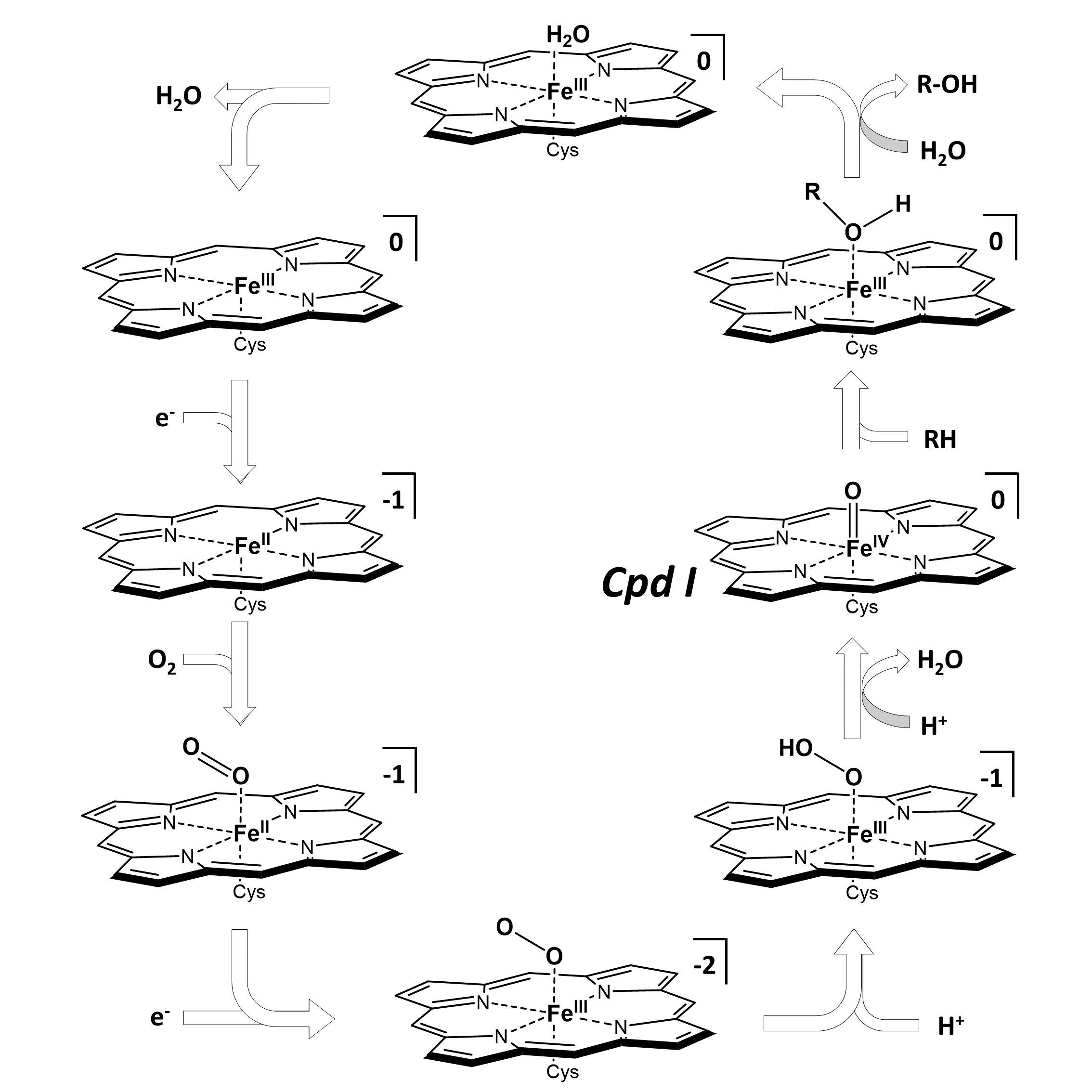}
\caption{\label{fig:catal_cycle} Catalytic cycle of P450 to oxidize aliphatic R-H bonds: Starting with water and Cys-thiolate bound heme (top) as resting state, water dissociates, and subsequently an electron is added to the system yield a 5-valent Fe(II) complex. Now molecular oxygen is associating to Fe(II) and stepwise addition of an electron and two protons cleaves the O-O bond to release one water entity. The remaining species is Compound I, where an oxygen atom is bound to Fe (formally in oxidation state IV). Compound I is the reactive species, and the addition of an aliphatic system (here denoted by RH) leads to an insertion of the oxygen into the R-H bond. The formed alcohol R-OH dissociates and water takes its place, yielding the resting state again and thereby closing the catalytic cycle.}
\end{figure}

The superfamily of cytochromes P450 (CYPs)  are membrane bound heme containing enzymes which function mostly as monooxygenases. In the human genome 57 CYP isoforms are encoded, moreover, it is the largest family of hemoproteins known, with probably more than 300.000 members throughout all organisms \cite{Nelson:2018}.  CYPs also catalyze other reactions like dealkylations, dehalogenations, and epoxidations \cite{Shaik:2010}. Still, the oxidation function is most important, being involved in the biosynthesis of steroids, hormones, as well as fatty acids \cite{Hajeyah:2020}. However, the major role of CYPs, which is the task they are known for, lies in the detoxification of organisms. The most common detoxification mechanism, displayed in the catalytic cycle in Figure~\ref{fig:catal_cycle}, involves a single oxygen insertion into C-H bonds of CYP substrates, thereby generating a hydroxy-group, which enables the further processing (like glucuronidation) or enhances direct excretion due to higher hydrophilicity of the metabolites. The oxidation by CYPs is also the most common metabolization mechanism  for drugs in humans, where more than 70\% of all drugs are metabolized by just two CYP isoforms (CYP 3A4 and CYP 2D6) \cite{Guengerich:2008}. Therefore, the interaction of CYPs and small organic molecules is of high interest in drug design. Structurally the CYPs contain a buried heme-iron center, where the iron is bound to a highly conserved cysteine thiolate side chain of the protein. The secondary and tertiary structure has been quite conserved through evolution \cite{Johnson:2013} and especially the cysteine facing amino acid environment close to the heme is highly conserved in CYPs. However, on the distal side of the heme group channels towards the solvent allow the entry of small organic molecules to the CYP active site above the heme group. In the case of CYP 3A4, which is known to roughly metabolize 50\% of all marketed drugs \cite{Liu:2007}, also more than one substrate molecule can be accommodated in the active site\cite{Boobis:2009}.

\begin{figure}[ht]
\centering
\includegraphics[width=0.95\textwidth]{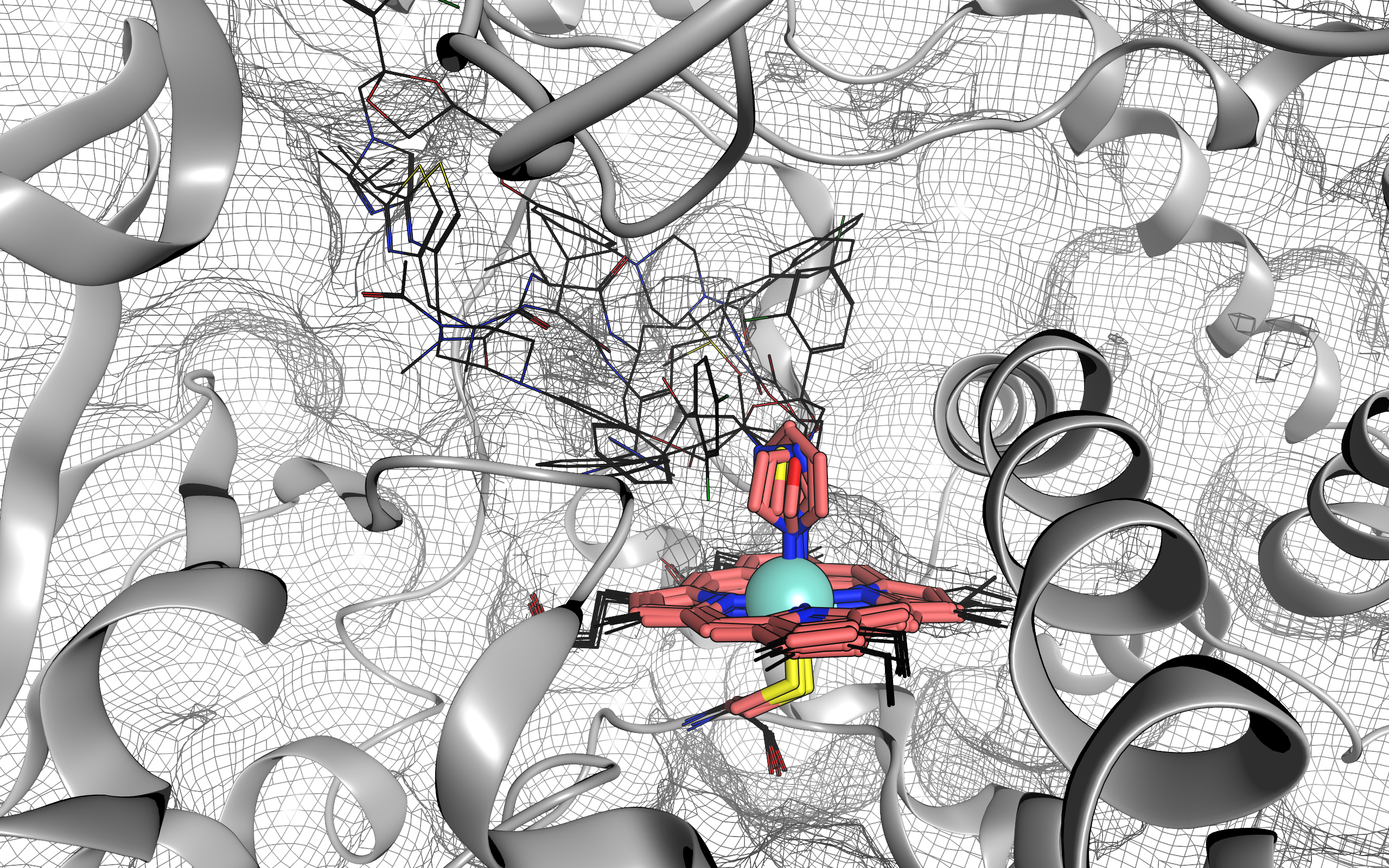}
\caption{Overlay of various CYP 3A4 structures bound to N-arene containing inhibitors - pdb structures 6MA7, 6MA6, 4I4G, 2V0M and 3NXU are displayed. Heme, thiolate, heteroarene-substructure of inhibiting drugs are displayed in salmon cylinder representation, the rest of the inhibitor structures as well as the thiolate bearing cysteine are shown as black line representations, and the rest of the CYP 3A4 protein is drawn as cartoon backbone and mesh surface in grey.
}
\label{fig:3a4}
\end{figure}

One has to distinguish various fundamentally different ways of interactions between small molecules and CYPs \cite{Zhou:2008}. In the most straightforward role of CYPs, as mentioned above, drugs act as substrates, which become metabolized and subsequently excreted. Another, more problematic way of interaction is, if small molecules act as inhibitors of CYPs. A CYP inhibitor binds to a certain isoform and hampers the proper catalytic functioning of the CYP, either by blocking the entry channel and/or the active site or through allosteric mechanisms \cite{Deodhar:2020}. In the case of mechanism based inhibition, a drug first behaves as a substrate, and its oxidized metabolite acts subsequently as an inhibitor of the respective CYP. Many drugs are known to inhibit some CYP isoforms \cite{Hakkola:2020}, and the potential problem of inhibitors lies in the potential of drug-drug interactions. The reason is that inhibition of drug metabolism results in reduced degradation of potential other substrates of that CYP isoform. In the worst case this leads to drug accumulation and higher drug levels in the body, increasing the risk for potential side effects \cite{Gajula:2021}. A variety of inhibitor bound CYP structures is known, and a reoccurring structural motif are aromatic nitrogen atoms that are directly coordinated to the iron center of the heme group \cite{Shah:2011}. In Figure~\ref{fig:3a4} we show a selection of inhibiting drug bound structures of CYP 3A4, where nitrogen arene substrucures of the inhibitors directly coordinate to the heme iron. CYP inhibiting substructures include triazole (contained in fluconazole, antifungal drug - pdb: 6MA7), pyridine (metyrapone, drug against adrenal insufficiency - pdb: 6MA6), oxazole (desoxyritonavir analogue - pdb: 4I4G), imidazole (ketoconazole, antifungal - pdb: 2V0M) and thiazole (ritonavir, co-medication to HIV treatment - pdb: 3NXU). From Figure~\ref{fig:3a4} a clear interaction pattern can be deduced, as the nitrogen atom of the heteroarene structures, directly binds to the iron center as 6\textsuperscript{th} coordination partner in addition to the porphyrine nitrogens and the thiolate. Inhibitors, therefore often directly interact with the iron and the interaction strength contributes strongly to the overall potency of an inhibitor.

Another important mechanism of interaction between CYPs and small molecules is CYP induction, where a drug changes the expression levels of CYPs. This is an indirect mechanism and mostly occurs through nuclear hormone receptors, also potentially impacting drug levels \cite{Lin:2012}. Whereas moderate drug metabolism by CYPs is beneficial for enabling drug excretion, usually CYP inhibition and CYP induction are mechanisms, which are usually avoided. Therefore a proper understanding of the molecular actions of small ligands on P450 enzymes as well as an understanding of the catalytic mechanism is of central importance for drug design. The heme group with its iron center, however, causes difficulties in the theoretical description of the relevant complexes, due to strong dependence on the protein environment and the high degree of electron correlation in the catalytic center. 

\subsubsection{Geometries}\label{app:geometries_from_xray}
The geometries of the model systems were derived from experimental X-ray structures of CYP3A4 and are displayed in Figure~\ref{fig:model}. The PDB access codes for the structures are 1TQN and 6MA6 for the water- and pyridine-bound complexes respectively. The model systems were prepared from the crystal structure by removal of all non-iron coordinating entities of the protein and the solvent. The iron-coordinating cysteine is truncated to methyl-thiolate and all porphyrine-attached substituents have been removed. The coordinating 6$^{th}$ ligand of the iron was truncated at the pyridine in the case of 6MA6. The pentacoordinated state has been modeled by removing the water from the 1TQN model system. Cpd~I models were derived from the water bound model system by removing both water bound hydrogens. 
The prepared geometries were optimized by UB3LYP~\cite{Vosko1980-uw, Lee1988-pi, Becke1993-nv,Stephens1994-cm} with the cc-pVDZ basis set~\cite{Dunning1989} employing LANL2DZ pseudopotentials~\cite{Wadt1985-uw, Hay1985-ma, Hay1985-eo} on the iron center by Gaussian09~\cite{g09} and standard convergence criteria. To improve convergence, the virtual orbitals were shifted by 0.1~Hartree. All model systems were of neutral charge and the overall spin multiplicity was set to 2 and 6 for low and high spin complexes respectively. 

\section{Computational details for full space and active space calculations}\label{app:software_implementation_details}
\subsubsection{Full valence space calculations}
To estimate the degree to which these compounds are strongly correlated, $\kappa$-OOMP2 and CCSD calculations were performed on these systems. Although these methods are single-reference methods, they were chosen to provide a reasonable estimate of the total electronic structure, as well as provide probes (such as the $\max(|t_1|)$ and $\max(|t_2|)$ diagnostics from CCSD) for evaluating potential multireference character. A complete discussion of the multireference character metrics used in this work can be found in Appendix~\ref{app:muliref_stats}.  The $\kappa$-OOMP2 calculations were performed in the full electronic space, while for efficiency CCSD utilized the common frozen-core approximation. These calculations were performed using the Q-Chem 5.4 electronic structure package~\cite{epifanovsky2021}. A value of $\kappa = 1.1$ was used for the $\kappa$-OOMP2 method, as recommended for transition metal complexes~\cite{shee2021regularized}. The geometry for each molecule corresponded to the B3LYP high-spin (S=5/2) optimized geometry. All calculations used the cc-pVDZ~\cite{Dunning1989} basis with the RI approximation. To test the impact of reference orbitals, CCSD calculations were performed using UHF, UB3LYP~\cite{Vosko1980-uw, Lee1988-pi, Becke1993-nv,Stephens1994-cm}, and UBP86~\cite{Perdew1986-vu, Becke1988-fc} orbitals. B3LYP and BP86 functionals were chosen to compare with a hybrid and a GGA functional, and the two functionals are among the most commonly used in chemistry. All references corresponded to stable, local minima. The orbitals obtained from KS-DFT were semi-canonicalized prior to the CCSD calculations. To evaluate the presence of multireference character in these compounds, the spin contamination of the reference $\langle S^2 - S_z^2 - S_z\rangle$, as well as the the $\max (|t_1|)$ and the $\max (|t_2|)$ diagnostics from CCSD were evaluated, where $t_1$ and $t_2$ are the CCSD amplitudes. From the $\kappa$-OOMP2 calculations, the natural orbital occupation numbers (NOONs) were computed by diagonalizing the corresponding one-particle density matrix (1PDM) in order to observe deviation from the idealized open-shell character~\cite{lee2019distinguishing}.
\subsection{Active space calculations}
To evaluate the quality of a given active space for fault-tolerant calculations on a quantum computer, DMRG calculations were performed on each active space for the doublet, quartet, and sextet multiplicities. DMRG calculations were performed with StackBlock (Block 1.5.0)~\cite{sharma2012} via the PySCF~\cite{sun2018, sun2020} interface. The reported energies are extrapolated based on the discarded weight computed from a reverse sweep schedule as described by  Olivares-Amaya {\it et al}~\cite{olivares-amaya2015}. $N$-electron valence state perturbation theory (NEVPT2)~\cite{Angeli2001} calculations as described by Sharma {\it et al.}~\cite{Sharma2017} were used to compute an estimate of dynamic correlation effects. 

In addition to the DMRG calculations, CCSD with non-iterative triples (CCSD(T))~\cite{Raghavachari1989} calculations were performed on the active space models with varying spin states. The coupled cluster calculations were performed using the PySCF~\cite{sun2018,sun2020} unrestricted coupled cluster code starting from the localized orbitals obtained from restricted open-shell Hartree-Fock, as detailed above. To account for changes in spin multiplicity, the orbitals were re-optimized within the active space prior to the CCSD(T) calculations.

\section{Details of DMRG calculations on Cpd I}\label{app:dmrg}
Some of the details of the DMRG calculations on our model of Cpd I are shown in Table~\ref{tab:dmrg_details}. The extrapolation error is estimated by dividing the difference between the largest calculation and the extrapolated value by 5. This empirical rule has been used in the past to provide a rough estimate of the extrapolation error~\cite{olivares-amaya2015}.
\begin{table}[h]
\centering
\begin{minipage}{0.8\textwidth}
\caption{Number of orbitals, max bond dimension, and extrapolation error for the DMRG calculations on Cpd I. Note that calculations on active spaces A-C where performed exactly, and the X active space was too large for practical calculations.}
\label{tab:dmrg_details}
\begin{ruledtabular}
\begin{tabular}{cccc}
    name &  \# of orbitals &  Max bond dimension & Extrapolation error-d,q,s (kcal/mol)\\ \hline
    B & 8& -& -\\
    C & 15& -& -\\
    D & 23& 1000& 2E-3, 2E-3, 7E-5\\
    E & 31& 1500& 1E-2, 1E-2, 6E-3\\
    F & 41& 1500& 3E-2, 6E-3, 2E-3\\
    G & 43& 1500& 1E-1, 2E-1, 1E-1\\
    X & 58& -& -\\ 
\end{tabular}
\end{ruledtabular}
\end{minipage}
\end{table}

\begin{figure}[H]
\centering
\includegraphics[width=0.3\textwidth]{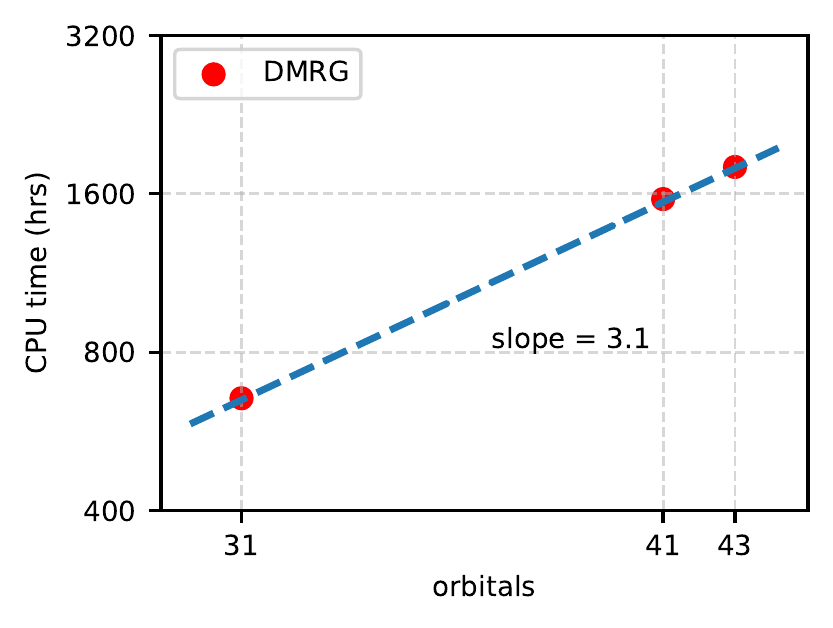}
\caption{\label{fig:cpd1_time} CPU time for DMRG calculations of fixed bond dimension ($M = 1500$) as a function of the active space size. The data are plotted on a log-log scale, and the slope of trendline, $\sim 3.1$ in this case, provides an estimate of the scaling for this fixed bond dimension. In practice, the high scaling of DMRG calculations comes from the increase in the necessary bond dimension with system size.}
\end{figure}

We also note that, for the purposes of resource estimation, the expected asymptotic scaling of $O(k^3)$ for $k$ active orbitals and a fixed bond dimension is empirically observed for the system sizes considered here (See Fig. ~\ref{fig:cpd1_time}).

\section{DMRG+NEVPT2 calculations on Cpd I}\label{app:nevpt2}
\begin{figure}[H]
\centering
\includegraphics[width=0.7\textwidth]{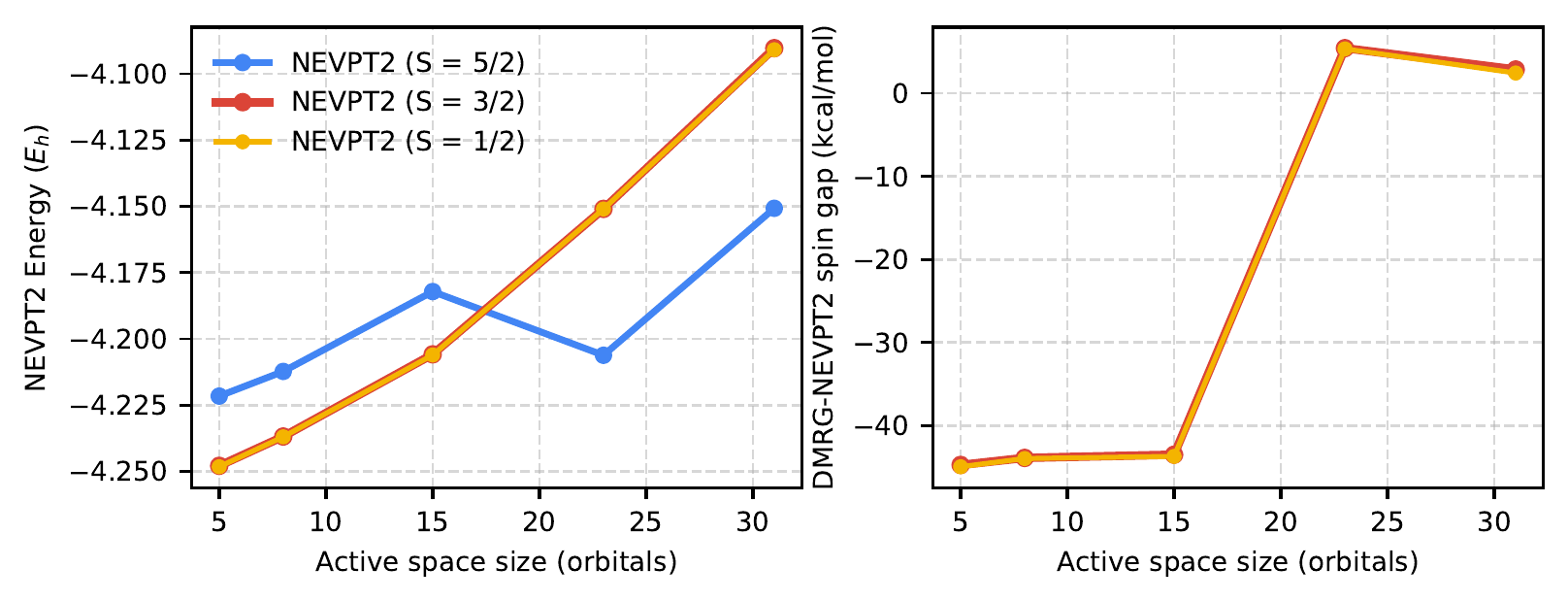}
\caption{\label{fig:cmp1_NEVPT2} \emph{left}: NEVPT2 energies, and, \emph{right}: DMRG-NEVPT2 spin gaps in active spaces A-E. In the D and E active spaces, the addition of the NEVPT2 correlation energy dramatically changes the spin ordering so that the high-spin state is lowest in energy.}
\end{figure}

In Figure~\ref{fig:cmp1_NEVPT2} we plot the NEVPT2 correction to each spin state for CASCI/DMRG wavefunctions in some of the smaller active spaces. In the D and E active spaces, small differences in the NEVPT2 correlation energy are nonetheless large enough to qualitatively change the spin ordering. However, this could be an artifact the NEVPT2 approximation itself, the approximate NEVPT2 implementation used in this work~\cite{Sharma2017}, or the approximate DMRG active-space wavefunction.

\section{Tensor Hypercontraction by Optimized CP3}\label{app:thc_cp3}
The Tensor Hypercontraction (THC) technique~\cite{hohenstein2012tensor,hohenstein2012communication,parrish2012tensor} has emerged as one of the most performant strategies for implementing qubitization based simulation strategies within fault tolerant quantum computing~\cite{PRXQuantum.2.030305}.  The THC decomposition is used in quantum chemistry to decompose the two-electron integral tensor from a four-tensor into five smaller tensors.  Mathematically, the THC is
\begin{align}
(ij|kl) = \sum_{PQ}^{M}X_{iP}X_{iQ}Z_{PQ}X_{kQ}X_{lQ}
\end{align}
where the THC rank $M$ scales as $M=\mathcal{O}(N \mathrm{poly}\log(1/\epsilon_{\mathrm{THC}}))$ where $\epsilon_{\mathrm{THC}}$ is the energy error due to the factorization per atom as we scale to the basis set limit~\cite{matthews2020improved}.  The THC factorization is related to the canonical polyadic (CP) decomposition~\cite{schutski2017tensor, pierce2021robust} but cannot be obtained through simple linear algebra relations. One strategy to obtain the THC factors is through the CP decomposition of geminal terms through alternating least-squares~\cite{schutski2017tensor} with an iteration complexity of $\mathcal{O}(r^{4}M$) where $r$ is the size of the single particle basis.  Extensive numerical studies suggest the rank of $M$ should scale linearly in the basis set size $r$ with a small integer prefactor (usually 3-6).  This has been benchmarked across a wide variety of chemical applications. Our interest in the THC factorization is that the cost of the repeated circuit primitives for the qubitized quantum walk oracles \textsc{SELECT} and \textsc{PREPARE} can be directly related to the $\mathrm{L}_{1}$-norm of the central tensor $Z_{PQ}$.  Thus our goal is not only to perform a high accuracy THC decomposition on the two-electron integral four-tensor but also minimize the norm of the central tensor. The norm minimization is not directly taken into consideration through standard THC decomposition techniques introduced in the quantum chemistry field. 

Following Ref.~\cite{PRXQuantum.2.030305} we start with an initial guess for the tensors $X_{iP}$ and $Z_{PQ}$ and optimize the $\mathrm{L}_{2}$ difference loss function
\begin{align}\label{eq:thc_loss}
L(X,Z) = \sum_{ijk}|(ij|kl) - \sum_{PQ}^{M}X_{iP}X_{iQ}Z_{PQ}X_{kQ}X_{lQ}|^{2} + C \sum_{PQ}|Z_{PQ}|
\end{align}
by L-BFGS-B implemented in Scipy. Due to the non-linear nature of the objective function the quality of the initial starting guess can drastically change the likelihood of finding a stationary point corresponding to a factorization that minimizes the $\mathrm{L}_{2}$ loss and has a small $L_{1}$\--norm for the central tensor. To find an initial $X$ and $Z$ we use the symmetric CP decomposition, denoted CP3, on the Cholesky factors, $B$, of the 2-electron integral tensor  
\begin{align}
(ij|kl) = \sum_{\chi}B_{ij, \chi}B_{kl, \chi} = \mathbf{B}\mathbf{B}^{T}
\end{align}
where $\mathbf{B}$ is obtained either from an SVD or improved scaling algorithms~\cite{epifanovsky2013general}.  Forming a symmetric decomposition of $\mathrm{B}$
\begin{align}
B_{ij,\chi} = \sum_{\tau}\beta_{i,\tau}\beta_{j,\tau}\zeta_{\chi,\tau}
\end{align}
is accomplished with alternating least-squares implemented in the C++ Basic Tensor
Algebra Subroutines (BTAS) library~\cite{BTAS}.  A Python implementation of the cost function and gradients for Eq.~\eqref{eq:thc_loss} was used to optimize the THC tensors for all systems we considered.  The $\mathrm{L}_{1}$ regularizer $C$ in Eq.~\eqref{eq:thc_loss} was set such that the $\mathrm{L}_{2}$ difference had the same strength as the $L_{1}$ part of the loss function given an initial set of THC factors from CP3.  We found that without regularization of the $\mathrm{L}_{1}$ norm THC decomposition from CP3 followed by L-BFGS-B optimization can provide erratically large $|Z_{PQ}|$ leading to erratic fault tolerant cost estimates for simulation of the systems considered here. As a consistency check we ensure that as the THC rank $M$ is increased the $\mathrm{L}_{2}$ norm of the difference between the true tensor and THC reconstructed two-electron integral tensor decreases along with a smooth increase in $\sum_{PQ}|Z_{PQ}|$.  The smooth convergence of $\lambda$, CCSD(T) error, and logical qubit count for different size THC factors is shown in Table~\ref{tab:cpd1_thc_convergence}.

\begin{figure}[H]
    \centering
    \includegraphics{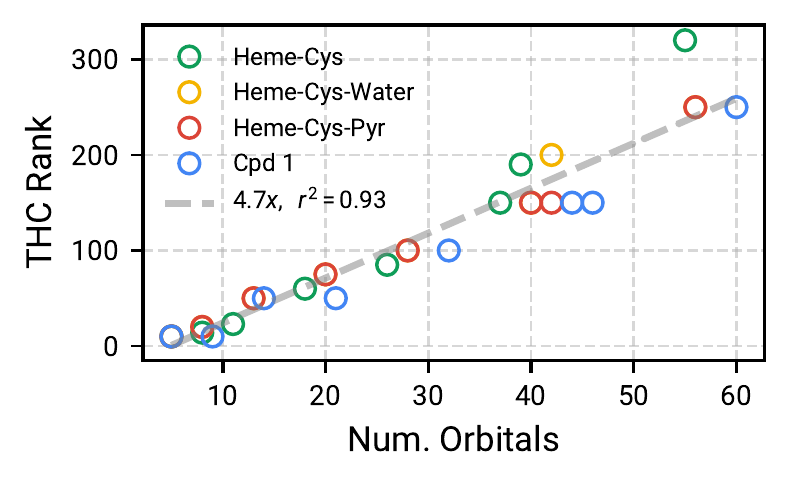}
    \caption{Number of orbitals versus the THC rank for active spaces A through X for all four compounds.}
    \label{fig:num_orbs_vs_thc_rank}
\end{figure}

To minimize pre-computation for quantum algorithms it is important to select the correct rank $M$ for THC without having to search.  In Figure~\ref{fig:num_orbs_vs_thc_rank} we plot the THC rank obtained from searching for a CCSD(T) error less than 1 milliHartree with respect to the full two-electron integral tensor.  We observed that the THC rank is approximately 5 times the size of the one-particle basis used in the problem which agrees well with previous chemistry studies.
\section{Active space sizes}\label{app:active_space_orb_n}
Starting from the high spin ROHF orbitals localized with the Pipek-Mezey scheme a class of active spaces is selected for each compound in the high spin configuration. \textcolor{black}{The logic for generating the active space hierarchy is to identify the orbitals most relevant to the ligand binding and spin character, expanding outward in terms of relevance and/or spatial localization, all the while adding same character virtual orbitals as needed to maintain a constant (roughly 50\%) filling fraction. The process is as follows: the open-shells are expected to be most important for spin-character so the five open-shells are selected for the first active space.  After the Fe d-orbitals are expected to be most important so the remaining occupied 3d orbitals are selected to yield active space `B' along with Iron 4s and (spatially) nearby iron-nitrogen anti-bonding orbitals to yield a roughly 50\% filling fraction. Subsequently, axial ligand (anti-)bonding orbitals were added to this set to yield active space `C' along with Iron orbitals mixing 4p and d' orbitals. These axial orbitals were selected as they are expected to be the most relevant to both the ligand binding as well as the overall spin state character. Finally, we augment the active spaces by expanding to orbitals spatially outward from this ``core active space'' of metal and axial metal-ligand orbitals. Iron-nitrogen sigma orbitals are spatially closest, and adding these along with their corresponding virtuals yield active space `D'. The heme $\pi$ orbitals are expected to stabilize the metal core, so we next added the heme $\pi$ orbitals from the nitrogen, as these bond directly to the metal (active space `E'). Next, we expand to the heme $\pi$ orbitals from the carbon atoms (active space `F'). Finally, the active spaces were augmented with the more distal carbon-sulfur (anti-)bonding orbitals (active space `G') and then the less-important heme carbon-nitrogen sigma orbitals (active space `X').} The same set of orbitals were used for active space calculations at different spin-states.  
\begin{table}[H]
\centering
\begin{minipage}{0.6\textwidth}
    \caption{Active space systems with number of orbitals and number of electrons in the active space. `Rest' corresponds to the resting state of the CYP active site which involves a water molecule as the sixth coordination to the cys-Fe(P) complex. `Empty' corresponds to the pentacoordinate cys-Fe(P) system with no sixth ligand. `Cpd1' is compound 1 and inhibited is the cys-Fe(P) system with a pyridine inhibitor coordinate to the sixth ligand site.}
    \label{tab:active_space_size_in_app}
\begin{ruledtabular}
    \begin{tabular}{ccccccccc}
    Active Space & \multicolumn{2}{c}{rest} & \multicolumn{2}{c}{empty} & \multicolumn{2}{c}{inhibited} & \multicolumn{2}{c}{Cpd1} \\
                 & orb. & elec. &  orb. & elec.&  orb. & elec. &  orb. & elec. \\
    \hline
     A   &  5 &  5 &  5 &  5 &  5 &  5&  5 &  5\\ 
     B   &  8 &  9 &  8 &  9 &  9 & 11&  8 & 11\\
     C   & 13 & 15 & 11 & 13 &  9 & 17& 15 & 19\\
     D   & 20 & 23 & 18 & 21 & 21 & 25& 23 & 25\\
     E   & 28 & 31 & 26 & 29 & 32 & 35& 31 & 33\\
     F   & 40 & 43 & 37 & 39 & 44 & 49& 41 & 45\\
     G   & 42 & 45 & 39 & 41 & 46 & 51& 43 & 47\\
     X   & 56 & 61 & 55 & 57 & 60 & 67& 58 & 63\\
    \end{tabular}
\end{ruledtabular}
\end{minipage}
\end{table}
\end{document}